\documentclass[a4pape,useAMS,usenatbib]{mn2e}
\input{epsf}

\def\lsim{ \lower .75ex \hbox{$\sim$} \llap{\raise .27ex \hbox{$<$}} }
\def\gsim{ \lower .75ex \hbox{$\sim$} \llap{\raise .27ex \hbox{$>$}} }
\def\g{{\tt GALFORM}}

\begin{document}

\title[LRGs in hierarchical models]
{Luminous Red Galaxies in hierarchical cosmologies}

\author[Almeida et~al.]{
\parbox[t]{\textwidth}{
\vspace{-1.0cm}
C.\,Almeida$^{1}$,
C.\,M.\,Baugh$^{1}$,
D.\,A.\,Wake$^{1}$,
C.\,G.\,Lacey$^{1}$,
A.\,J.\,Benson$^{2}$,
R.\,G.\,Bower$^{1}$
K.\,Pimbblet$^{3}$.
}
\\
$^{1}$Institute for Computational Cosmology, Department of Physics, 
University of Durham, South Road, Durham, DH1 3LE, UK.\\ 
$^{2}$California Institute of Technology, MC 130-33, 1200 E. California 
Blvd, Pasadena, CA 91125, USA.\\ 
$^{3}$Department of Physics, University of Queensland, Brisbane, QLD 4072, 
Australia.\\
}
 
\maketitle
\begin{abstract}
Luminous red galaxies (LRGs) are much rarer and more massive 
than $L_*$ galaxies. Coupled with their extreme colours, LRGs 
therefore provide a demanding testing ground for the physics of massive 
galaxy formation. We present the first self-consistent predictions 
for the abundance and properties of LRGs in hierarchical structure 
formation models. We test two published models which use quite 
different mechanisms to suppress the formation of massive galaxies: 
the Bower et~al. (2006) model, which invokes ``AGN-feedback'' to prevent 
gas from cooling in massive haloes, and the Baugh et~al. (2005) model 
which relies upon a ``superwind'' to eject gas before it is turned 
into stars. Without adjusting any parameters, the Bower et~al. model 
gives an excellent match to the observed luminosity function of LRGs 
in the Sloan Digital Sky Survey (with a median redshift of $z=0.24$) 
and to their clustering; the Baugh et~al. model is less successful 
in these respects. Both models fail to match the observed abundance 
of LRGs at $z=0.5$ to better than a 
factor of $\approx 2$. In the models, LRGs are 
typically bulge dominated systems with stellar masses of 
$\approx 2 \times 10^{11}h^{-1}M_{\odot}$ 
and velocity dispersions of $\sigma \sim 250 {\rm km s}^{-1}$. 
Around half of the stellar mass in the model LRGs is already formed 
by $z \sim 2.2$ and is assembled into one main progenitor 
by $z \sim 1.5$; on average, only 25\% of the mass of the main 
progenitor is added after $z \sim 1$. LRGs are predicted to be 
found in a wide range of halo masses, a conclusion which relies 
on properly taking into account the scatter 
in the formation histories of haloes. Remarkably, we find that the 
correlation function of LRGs is predicted to be a power law down to small 
pair separations, in excellent agreement with observational estimates.
Neither the Bower et~al. nor the Baugh et~al. model is able to reproduce 
the observed radii of LRGs. 
\end{abstract}

\section{Introduction}

Over the past few years, the most rapidly developing aspect of 
galaxy formation modelling has been the formation of massive 
galaxies (see Baugh 2006 for a review). On employing the standard 
White \& Frenk (1991) model for the radiative cooling of gas in 
massive dark matter haloes, hierarchical models have tended to 
overproduce luminous galaxies. One pragmatic solution to this problem 
is to simply stop ``by hand'' the formation of stars from cooling flows  
in high circular velocity haloes (Kauffmann et~al. 1993). A variety of 
physical mechanisms have been proposed to account for the suppression 
of the star formation rate in massive haloes, including: (i) the injection 
of energy into the hot gas halo to reduce its density and hence increase the 
cooling time (Bower et~al. 2001; McCarthy et~al. 2007); (ii) the fragmentation 
of the hot halo in a multi-phase cooling model (Maller \& Bullock 2004); 
(iii) the complete ejection of gas from the halo in a ``super wind'' (Benson et~al. 2003); 
(iv) the suppression of the cooling flow due to heating by an AGN 
(Croton et~al. 2006; Bower et~al. 2006); (v) 
thermal conduction of energy within the hot halo 
(Fabian et~al. 2002; Benson et~al. 2003). With such a range of possible 
physical processes to choose from, it is important to develop tests of 
the models which can distinguish between them. 
The mechanisms invoked to suppress the formation of bright galaxies 
could scale in different ways with redshift, leading to different 
predictions for the galaxy properties at intermediate and high redshift.   

In this paper we present new tests of the physical processes invoked to 
suppress the formation of bright galaxies. At the present day, the bright 
end of the luminosity function is dominated by early-type galaxies with 
passively evolving stellar populations (e.g. Norberg et~al. 2002). Here we focus 
on a subset of bright galaxies, luminous red galaxies (LRGs) and test 
the model predictions for the abundance and properties of these red, massive galaxies. 
LRGs were originally selected from the Sloan Digital Sky Survey (SDSS, York et~al. 2000) 
on the basis of their colours and luminosities (Eisenstein et~al. 2001). The red colour 
selection isolates galaxies with a strong 4000 \AA \,\, break and a passively 
evolving stellar population. The galaxies selected tend to be significantly 
brighter than $L_*$. The SDSS sample has a median redshift of $z \sim 0.3$. 
Recently, the construction of LRG samples has been extended to higher 
redshifts, using SDSS photometry and the 2dF and AAOmega spectrographs 
($z \sim 0.5 $, Canon et~al. 2006; $z \sim 0.7$, Ross et~al. 2007b). 
On matching the colour selection between the SDSS and 2SLAQ surveys, 
the evolution in the luminosity function of LRGs between $z \sim 0.3 $ and 
$z \sim 0.5$ is consistent with that expected for a passively evolving stellar 
population (Wake et~al. 2006). This has implications for 
the stellar mass assembly of the LRGs, with the bulk of the stellar mass 
appearing to have been in place a significant period before the LRGs are observed 
(Wake et~al. 2006; Roseboom et~al. 2006; Brown et~al. 2007). Due to their strong clustering 
amplitude and low space density, LRGs are efficient probes of the large-scale structure 
of the Universe. The clustering of LRGs has been exploited to constrain 
cosmological parameters (e.g. Eisenstein et~al. 2005; H\"{u}tsi 2006; 
Padmanabhan et~al. 2007). The clustering of LRGs on smaller scales has been 
used to constrain the mass of the dark matter haloes which host these galaxies 
and to probe their merger history (Zehavi et~al. 2005; Masjedi et~al. 2006;  
Ross et~al. 2007a). 

To date, surprisingly little theoretical work has been carried out to see if 
LRGs can be accommodated in hierarchical cosmologies, and only very simple 
models have been used. Granato et~al. (2004) considered a model for the 
formation of spheroids in which the quasar phase of AGN activity 
suppresses star formation in massive galaxies, and found reasonable 
agreement with the observed counts of red galaxies at high redshift. 
Hopkins et~al. (2006) used the observed luminosity function of 
quasars along with a model for the lifetime of the quasar phase suggested by 
their numerical simulations to infer the formation history of spheroids, and 
hence red galaxies. Conroy, Ho \& White (2007) use N-body simulations to study 
the merger histories of the dark matter haloes that they assume host LRGs. 
Using a simple model to assign galaxies to progenitor haloes, they argue that 
mergers of LRGs must be very efficient, or that LRGs are tidally disrupted, 
in order to avoid populating cluster-sized haloes with too many LRGs. 
Barber et~al. (2007) used population synthesis models coupled with 
assumptions about the star formation histories of LRGs to infer the age 
and metallicity of their stellar populations. 

Here we present the first fully consistent predictions for LRGs from hierarchical 
galaxy formation models, using two published models, namely Baugh et~al. (2005) and 
Bower et~al. (2006). These models, both based on the {\tt GALFORM} semi-analytical 
code (Cole et~al. 2000), carry out an {\it ab initio} calculation of the fate of 
baryons in a cold dark matter universe. The models predict the star formation 
and merger histories for the whole of the galaxy population, producing broad-band 
magnitudes in any specified pass band. Hence, LRGs can be selected from the model 
output using the same colour and luminosity criteria that are applied to the real 
observational data. The models naturally predict which dark matter haloes host  
LRGs. As we will see later, a key element in shaping the ``halo occupation 
distribution'' of LRGs is the scatter in the merger histories of dark matter 
haloes, which has been ignored in previous analyses. 

This paper extends the work of Almeida et~al. (2007) in which we tested the 
predictions of the same two galaxy formation models for the scaling relations 
of spheroids, such as the radius-luminosity relation and the fundamental plane. 
The galaxies we consider in this paper represent a much more extreme 
population than those studied in our previous work. LRGs are much rarer and 
significantly brighter than $L_{*}$ galaxies and even than the early-type population 
as a whole. In this paper we concentrate on massive red galaxies at low and intermediate 
redshifts where large observational samples exist; in a companion study, 
we test the model predictions for ``extremely red objects'' at high redshift (Gonz\'alez-P\'erez et~al., 
in preparation). 
We remind the reader of the key features of the two models in Section 2 
(see also the comparison given in Almeida et~al. 2007). In Section 3, we explain the 
selection of LRGs and show some basic predictions for the abundance and properties of LRGs.
In Section 4, we present predictions for the clustering of LRGs and in Section 
5 we examine how the stellar mass of LRGs is built up in the models. 
Our conclusions are given in Section 6. 

\section{Modelling the galaxy population}
\label{section:model}

In this section, we give a brief outline of the two versions 
of the \g~model which we study in this paper, the Baugh et~al. 
(2005) and the Bower et~al. (2006) models. An introduction to 
the semi-analytical approach to modelling the formation of 
galaxies can be found in the review by Baugh (2006). The \g~model 
itself is described in detail by Cole et~al. (2000). The 
superwind feedback model used by Baugh et~al. was introduced 
by Benson et~al. (2003) and is also discussed by Nagashima 
et~al. (2005a). 

Both the Baugh et~al. and Bower et~al. models are calibrated 
against a subset of the observational data available for local 
galaxies. The Bower et~al. model gives a somewhat better match 
to the sharpness of the break in the optical and near-infrared 
galaxy luminosity functions than the Baugh et~al. model.
Other outputs from the models besides these local calibrating data are model 
predictions. The Bower et~al. model also gives an excellent match 
to the evolution of the stellar mass function inferred from 
observations. The Baugh et~al. model has been tested extensively 
and reproduces a wide range of datasets: the number counts and redshift 
distribution of galaxies in the sub-mm and the luminosity function of 
Lyman-break galaxies (Baugh et~al. 2005), the mid-infrared luminosity 
functions as measured using Spitzer (Lacey et~al. 2008), the metal 
content of the intracluster medium (Nagashima et~al. 2005a), 
the metallicity of elliptical galaxies (Nagashima et~al. 2005b), 
the abundance of Lyman-alpha emitters 
and their properties (Le Delliou et~al. 2005; 2006) and some of 
the scaling relations of elliptical galaxies, including the 
fundamental plane (Almeida et~al. 2007). 

We emphasize that in this paper we do not vary any of the parameters 
which specify the Baugh et~al. and Bower et~al. models. Our goal is to 
expand the tests of the published models to include the predictions 
for the abundance and properties of luminous red galaxies. None of the 
datasets originally used to set the model parameters had any explicit 
connection to bright red galaxies at the redshifts of interest in this 
paper. The results we present are therefore genuine predictions of the 
model and represent a powerful, ``blind'' test of the semi-analytical 
methodology. A key constraint in setting the model parameters is 
the requirement that they reproduce as closely as possible the bright 
end of the present day galaxy luminosity function, which is dominated 
by galaxies with red colours and passive stellar populations (e.g. 
Norberg et~al. 2002). Matching the observed properties of LRGs 
therefore acts as a test of the evolution of the bright end of 
the luminosity function in the models, as traced by galaxies 
with passive stellar populations.  

A full description of the two models is, of course, given in the 
original papers. A comparison of the ingredients in the models can 
be found in Almeida et~al. (2007; see also Lacey et~al. 2008). Here, 
for completeness, we give a brief summary of where the principal 
differences lie between the models:

\begin{itemize} 
\item {\it Dark matter halo merger trees.} 
The Baugh et~al. model employs halo merger trees generated using 
the Monte Carlo algorithm introduced by Cole et~al. (2000). In the 
Bower et~al. model, the halo merger histories are extracted from the 
Millennium Simulation \citep{springel_2005}. A comparison of the 
predictions of galaxy formation models made using these two approaches 
to produce merger trees shows that they yield similar 
results for galaxies brighter than a 
threshold magnitude which is set by the mass resolution of the 
N-body trees \citep{helly}. In the case of the Millennium merger trees, 
this limit is several magnitudes fainter than $L_{*}$ and so has no 
impact on the results presented in this paper, which concern much 
brighter galaxies.

\item{\it Feedback processes.}
Both models use the ``standard'' supernova driven feedback common 
to essentially all semi-analytical models (though with different 
values for the parameters). In this scenario, supernovae and 
stellar winds reheat cooled gas and thus regulate the supply of gas 
available for subsequent star formation. The models differ in how 
they treat the reheated gas. In the Baugh et~al. model, the 
reheated gas is not considered as being available to cool again 
from the hot halo until the mass of the halo has doubled. At this 
point, the gas heated by supernova feedback is added to the hot gas 
halo of the new dark matter halo. Bower et~al., on the other hand, 
incorporate the reheated gas into the hot halo after a delay which is a 
multiple of the dynamical time of the dark matter halo. 

The two models use different feedback mechanisms to counter the 
overproduction of bright galaxies, which was a long-standing problem 
for hierarchical models (see Baugh 2006). Baugh et~al. invoke a wind 
which ejects cold gas from galaxies at a rate which is a multiple 
of the star formation rate \citep[see][]{b03}. The gas thus ejected 
is not allowed to recool, even in more massive haloes. 
This is another ``channel'' for the energy released by supernovae to 
couple to the cold gas reservoir available for star formation which 
operates alongside the feedback mechanism described in the previous 
paragraph. The superwind and the ``standard'' supernova feedback have 
distinct parameterizations in terms of the star formation rate, and 
differ in the fate of the reheated gas, as discussed above (see Lacey 
et~al. 2008 for an expanded discussion and for the respective equations). 
There is observational evidence for superwind outflows in the 
spectra of Lyman-break galaxies and in local starburst 
galaxies \citep{adelberger,wilman}. 
In the Bower et~al. model, the cooling of gas is suppressed in massive 
haloes due to the heating of the halo gas by the energy released by 
the accretion of matter onto a central supermassive black hole. 
The growth of the black hole is based on the model described 
by \citet{rowena}. 

\item{\it Hot gas distribution} 
Both models adopt a density profile for the hot gas halo of the 
form $\rho \propto (r^2 + r^{2}_{\rm core})^{-1}$. In the Bower 
et al. model, $r_{\rm core}$ is kept fixed at 0.1 of the virial 
radius. In the case of Baugh et al., the core radius is initial 
set to be one third of the scale length of the dark matter density 
profile (Navarro, Frenk \& White 1997). The core radius evolves with 
time in this model, as it is recomputed when a new halo forms to take 
into account that the densest, lowest entropy gas has cooled 
preferentially from the central regions of the progenitor 
haloes (see Cole et~al. 2000).

\item{\it Star formation.} 
In both models, there are two modes of star formation, quiescent star 
formation, which occurs in galactic disks, and starbursts. 
Baugh et~al. adopt a quiescent star formation timescale which is 
independent of the dynamical time of the galaxy, unlike Bower 
et~al.. Hence, galactic disks tend to be gas rich at high redshift 
in the Baugh et~al. model, whereas they are gas poor in the Bower 
et~al. model; this means that starbursts triggered by galaxy mergers 
tend to be more intense in the Baugh et~al. model than in Bower et~al. 
The later model also allow bursts which are the result of a galactic 
disk becoming dynamically unstable. 

\item{\it Stellar Initial Mass Function (IMF).}
Both models adopt a standard solar neighbourhood IMF, the Kennicutt (1998) 
IMF, for quiescent star formation. Bower et~al. also use this IMF in 
starbursts, whereas Baugh et~al. invoke a top heavy IMF, which is 
the primary ingredient responsible for this model's successful 
reproduction of the number counts of sub-mm galaxies.  
The yield we adopt is consistent with the choice of IMF. 
The choice of a top-heavy IMF in starbursts is controversial, 
but has been tested successfully against the metal content of the 
intra-cluster medium (Nagashima et~al. 2005a) and the metallicity 
of elliptical galaxies (Nagashima et~al. 2005b).

\item{\it Cosmology.} Baugh et~al. use the canonical 
($\Lambda$CDM) parameters: matter density, $\Omega_{0}=0.3$, cosmological 
constant, $\Lambda_{0} = 0.7$, baryon density, $\Omega_{b}=0.04$, 
a normalization of density fluctuations given by $\sigma_{8}=0.93$ 
and a Hubble constant $h=0.7$ in units of 100 km s$^{-1}$ Mpc$^{-1}$. 
(Note in Baugh et~al., the value of $\sigma_{8}$ is reported as 0.9, when 
this should be $\sigma_{8}=0.93$.) 
Bower et~al. adopt the cosmological parameters of the Millennium 
simulation (Springel et~al. 2005), which are in better agreement 
with the latest constraints 
from measurement of the cosmic microwave background radiation and large 
scale galaxy clustering (e.g. Sanchez et~al. 2006): $\Omega_{0}=0.25$, 
$\Lambda_{0} = 0.75$, $\Omega_{b}=0.045$, $\sigma_{8}=0.9$ and $h=0.73$. 
\end{itemize}

\section{LRG selection and basic properties}

In this section, we present the predictions of the Baugh et~al. 
and Bower et~al. models for the basic properties of low and intermediate 
redshift luminous red galaxies (LRGs) and  compare these with 
observational results from the SDSS and 2SLAQ LRG samples. 
LRGs are a subset of the overall early-type population with extreme 
luminosities and colours, so it is essential to match their selection 
criteria as closely as possible in order to make a meaningful test of 
the model predictions. 
We begin by reviewing the colour and magnitude selection used in 
these surveys (\S 3.1), before moving on to examine the predictions 
for the abundance of LRGs (\S 3.2). This issue is dealt with in 
further detail in \S 4, in which we focus on the clustering of LRGs. 
In \S 3.3, we compare the model predictions for a range of LRG properties 
with observations. Finally, in \S 3.4, we discuss the physical reasons 
for the differences between the predictions of the two models. 

\subsection{Sample selection: SDSS and 2SLAQ LRGs}
\label{section:selection}

The basic aim of LRG surveys is to select intrinsically bright 
galaxies which have colours consistent with those expected for 
a passively evolving stellar population \citep{ei}. The selection 
criteria used in the SDSS LRG and 2dF-SDSS LRG and QSO (2SLAQ) 
surveys are targetted at different redshift intervals and pick 
up very different number densities of objects. Full descriptions 
of the design of the respective surveys can be found in \citet{ei} 
and \citet{cannon}.  

Below, for completeness, we give a summary of the colour 
and magnitude ranges which define the LRG samples. In the case 
of the observational samples, Petrosian magnitudes were used for 
apparent magnitude selection and SDSS model magnitudes were used 
for colour selection. The SDSS filter system is described in 
Fukugita et~al (1996). In the case of \g~galaxies, we use the 
total magnitude. We consider two output redshifts in the \g~models, 
chosen to be close to the median redshifts of the observational 
samples; $z=0.24$ to compare with SDSS LRGs and $z=0.50$ to 
match the 2SLAQ LRGs.

In the case of the SDSS, two combinations of the $g-r$ and $r-i$ 
colours are formed: 
\begin{eqnarray}
c_{\perp} &=& (r-i) - (g-r)/4 - 0.177 \\
c_{\parallel} &=& 0.7 (g-r) + 1.2 [(r-i)-0.177].
\end{eqnarray}
The following conditions are then applied to select LRGs: 
\begin{eqnarray}
r_{\rm petro} &<& 19.2  \\ 
r_{\rm petro} &<& 13.116 + c_{\parallel}/0.3 \\
|c_{\perp}|   &<& 0.2 
\end{eqnarray}
In the case of 2SLAQ, somewhat different colour combinations 
are used: 
\begin{eqnarray}
d_{\perp} &=& (r-i) - (g-r)/8 \\
d_{\parallel} &=& 0.7 (g-r) + 1.2[(r-i)-0.177].
\end{eqnarray}
The selection criteria applied in the case of 2SLAQ are: 
\begin{eqnarray}
17.5 &<& i\,<\, 19.8 \\
0.5  &<& g - r\, <\, 3 \\
r-i  &<& 2\\
d_{\perp} &>& 0.65\\
d_{\parallel} &>& 1.6
\end{eqnarray} 
The colour equations (Eqs. 1, 2 6, 7) and the conditions applied to them 
(Eqs. 5, 11, 12) are designed to locate galaxies with appreciable 4000 \AA\ 
breaks in the $(g-r)$ vs $(r-i)$ plane over the redshift intervals of the 
two surveys (see Eisenstein et~al 2001 and Cannon et~al. 2006 for further 
details).

As we have already commented, these two sets of selection criteria give 
quite different number densities of LRGs. Here we {\it do not} attempt 
to tune the selection to match objects in the 2SLAQ LRG sample with those 
from the SDSS LRG sample. This was done  by \citet{david}, whose motivation 
was to study the evolution of the LRG luminosity function. Our aim instead 
is to test the galaxy formation models, so trying to match the selection 
to pick out similar objects between the two redshifts is not necessary. 


\subsection{Luminosity Function}
\label{section:lf}

The luminosity function is the most basic description of any 
galaxy population and is arguably the key hurdle for a model 
of galaxy formation to negotiate before considering other 
predictions. It is important to bear in mind that LRGs represent 
only a small fraction of the galaxy population as a whole, as can 
be seen by comparing the integrated space densities quoted in Table~1 
with the abundance of $L_*$ galaxies, which is around an order of 
magnitude higher. Reproducing the abundance of such rare galaxies 
therefore represents a strong challenge for any theoretical model.

\begin{figure}
{\epsfxsize=8.truecm
\epsfbox[18 144 592 718]{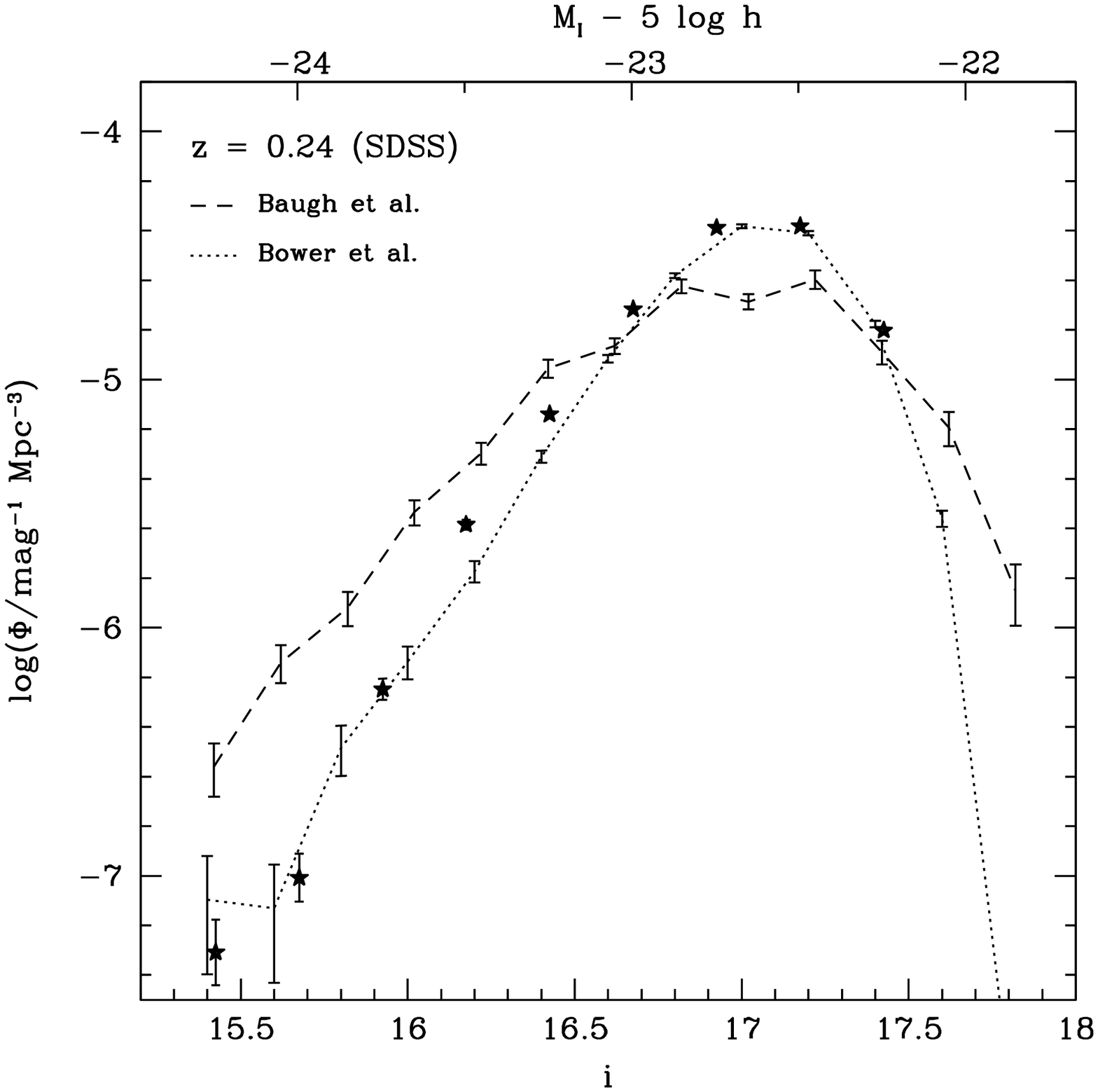}}
{\epsfxsize=8.truecm
\epsfbox[18 144 592 718]{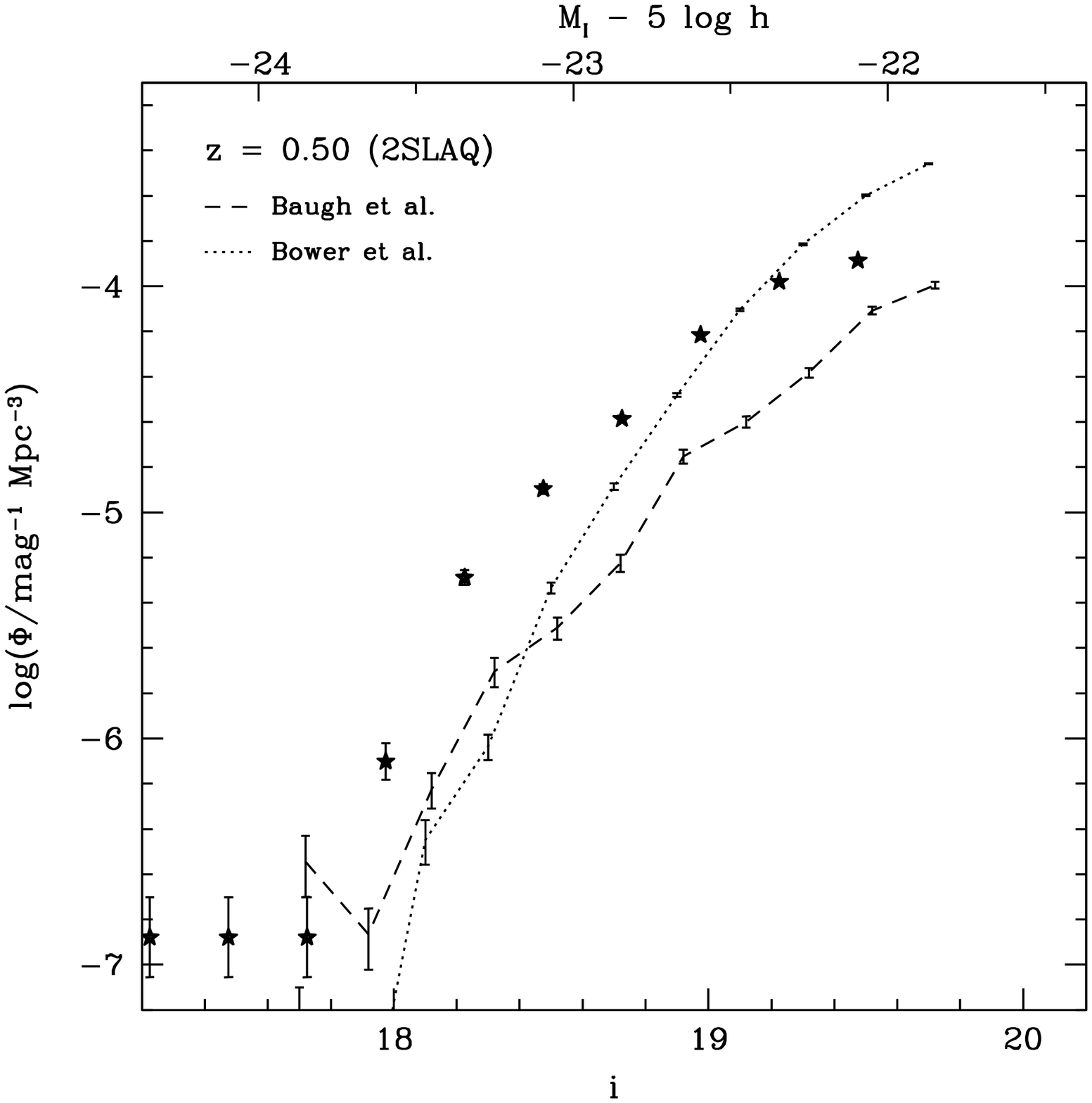}}
\caption{
The luminosity function of luminous red galaxies plotted 
as a function of apparent magnitude. 
The upper panel shows the 
SDSS LRG luminosity function at $z=0.24$, the median redshift of 
this sample, and the lower panel shows the results for the 2SLAQ 
sample at its median redshift, $z=0.5$. 
The upper axis labels show the absolute magnitude for reference 
(calculated from the apparent magnitude by subtracting the 
appropriate distance modulus for each panel). 
In each panel, the dashed line shows the number density 
of LRGs predicted by the Baugh et~al. model and the dotted 
line shows the prediction of the Bower et~al. model. 
The observational estimates are shown by the stars.
The error bars on the model predictions show the Poisson 
error due to the finite number of galaxies simulated. 
}
\label{fig:lfs}
\end{figure}

In Fig.~\ref{fig:lfs}, we compare the predictions of the \g~models 
for the luminosity function of LRGs with observational estimates.
This determination of the observed 2SLAQ and SDSS luminosity 
functions is different from that presented in Wake et~al. (2006). 
Here, we have estimated the observed luminosity functions in such 
a way as to minimize the corrections necessary to compare to the 
models. We have restricted both samples to tight redshift ranges around 
the model output redshifts, $0.22 < z < 0.26$ for the case of SDSS 
and $0.48 < z < 0.54$ for 2SLAQ. We then use simple K+e corrections 
derived from Bruzual \& Charlot (2003) stellar population synthesis models
(see Wake et~al. 2006) to correct the SDSS LRG magnitudes to z = 0.24 and
the 2SLAQ LRG magnitudes to z = 0.5. Since the redshift ranges considered 
here are so close to the target redshift, these corrections are 
very small, $< 0.01$ mags. The SDSS catalogue is then cut at 
$i^{0.24} < 17.5$ and the 2SLAQ catalogue is cut at $i^{0.5} < 19.6$. 
Both of these cuts are brighter than the magnitude limits of each survey 
and within the limited redshift ranges effectively produce volume-limited samples. 
The final samples contain 5217 and 2576 LRGs within 8.5x10$^{7} 
{\rm Mpc}^{3}$ and  3.9x10$^{7} {\rm Mpc}^{3}$ for SDSS and 2SLAQ
respectively (assuming $h=0.7=H_{0}/(100 {\rm km s}^{-1}{\rm Mpc}^{-1}$). 
Since the samples are approximately volume-limited it is trivial to 
produce the luminosity function including a correction for incompleteness 
in each survey (see Wake et~al. 2006). The integrated number densities of 
LRGs in the two surveys are listed in Table 1 after applying the respective 
completeness corrections.

In view of the fact that no model parameters have been adjusted in order 
to ``tune'' the predictions to better match observations of LRGs, both models 
come surprisingly close to matching the number density of LRGs in the SDSS 
sample, as Table~1 shows. In fact, the Baugh et~al. model slightly 
overpredicts the space density of LRGs at $z=0.24$ by 13\%, whereas 
the Bower et~al. model underpredicts only by 10\%. However, 
Fig.~\ref{fig:lfs} shows that the Baugh et~al. actually gives a poor match 
to the shape of the luminosity function, predicting too many bright LRGs. 
Although the difference looks dramatic on a logarithmic scale, 
the discrepancy has little impact on the integrated space density.

At the median redshift of the 2SLAQ sample, the comparison with the 
observational estimate of the luminosity function of LRGs is less impressive. 
The Baugh et~al. model now underpredicts the abundance of LRGs by a 
factor of $2$. Alternatively, the discrepancy is equivalent to 
a shift of about one magnitude in the $i$-band. 
The Bower et~al. model fares analogously, predicting 
around $100\%$ more LRGs than are seen in the 2SLAQ sample. 
This suggests that neither model is able to accurately follow the 
evolution of the luminosity function of very red galaxies over such 
a large lookback time (when the age of the universe is only around 60\% 
of its present day value). 
The abundance of LRGs is therefore quite sensitive to the way 
in which feedback processes are implemented in massive haloes. 

\begin{table}
\begin{center}
 \begin{tabular}{ccc}
 \hline
 Sample &   SDSS     & 2SLAQ \\
        & ($z = 0.24$) & ($z=0.50$)\\
        & ($10^{-5}$ Mpc$^{-3}$) & ($10^{-5}$ Mpc$^{-3}$) \\
 \hline
 Observed space density & 3.30 & 8.56 \\
 Baugh et~al. prediction & 3.74 & 4.75 \\
 Bower et~al. prediction & 2.99 & 18.11 \\
 \hline
 \end{tabular}
  \caption{
The space density of LRGs in the SDSS and 2SLAQ samples estimated 
at their median redshifts, compared with the predictions of the 
Baugh et~al. and Bower et~al. models. The number density in the 
table are quoted in units of $10^{-5}$ Mpc$^{-3}$. (The relevant 
$h$ is used for each model; for the data, $h=0.7$ is assumed.)
}
 \label{table:density}
 \end{center}
\end{table}


\subsection{Properties of LRGs}

In this section, we present a range of predictions for 
the properties of galaxies which satisfy the LRG selection 
criteria defined in section~\ref{section:selection}, for both 
the Baugh et~al. and Bower et~al. models, comparing with 
observational results whenever possible. We remind the 
reader that the models do not reproduce exactly the shape 
and normalization of the observed luminosity function of LRGs as 
seen in S 3.2, but instead bracket the observed abundances. 
Rather than perturb the selection criteria applied the model 
galaxies to better match the observed abundances, we have 
retained the full LRG selection criteria (\S 3.1) so that the 
model galaxies have the same colours and magnitudes as observed 
LRGs.  

\subsubsection{Stellar mass} 
\label{sec:mstar}

\begin{figure}
{\epsfxsize=8.truecm
\epsfbox[18 144 592 718]{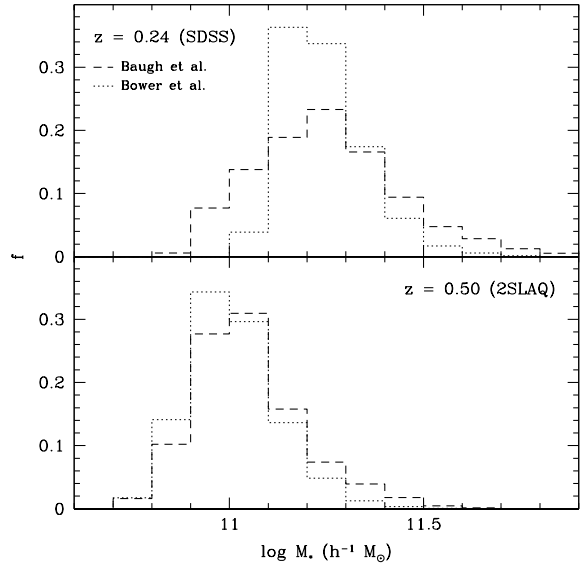}}
\caption{
The predicted stellar masses of model luminous red galaxies at 
$z=0.24$ (upper panel) and $z=0.50$ (lower panel). 
The predictions of the Baugh et~al. model are shown by the dashed histograms
and those of the Bower et~al. model by the dotted histograms. 
The distributions are normalized to give $\sum_i f_i = 1$.}
\label{fig:stellar.mass}
\end{figure}

The predicted stellar masses of LRGs are plotted in 
Fig.~\ref{fig:stellar.mass}. As expected from the high luminosities of 
LRGs, these galaxies exhibit large stellar masses. 
At $z=0.24$ the stellar masses range from $\sim 1\times10^{11}$ to 
$5\times10^{11}$ h$^{-1}$M$_{\odot}$, with a median of  
$1.7 \times10^{11}$ h$^{-1}$M$_{\odot}$. At $z=0.50$, 
the distribution shifts to lower stellar masses, with a median value  
of $\sim 1\times 10^{11}$ h$^{-1}$M$_{\odot}$.
The scatter in the distribution of stellar masses predicted by the 
Baugh et~al. model is somewhat larger than that in the Bower et~al. model 
at $z=0.24$. 
The median stellar mass 
is a property for which the two models agree closely, indicating 
that stellar mass is a robust prediction which is fairly insensitive 
to the details of the implementation of the physics of galaxy formation. 
The difference in the selection criteria applied to the two samples is 
responsible for picking up objects of quite different stellar 
masses. We shall see in subsequent comparisons that this basic difference 
between the LRG samples is responsible for differences in other model 
predictions. 

\subsubsection{Morphological mix} 
\label{sec:morph}

\begin{figure}
{\epsfxsize=8.truecm
\epsfbox[18 144 592 718]{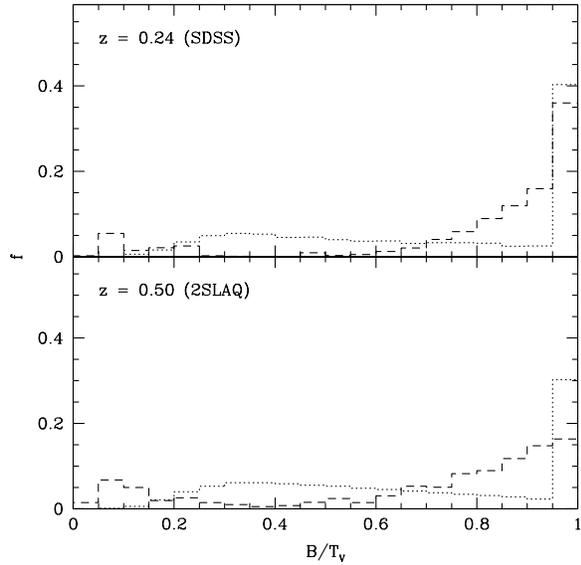}}
\caption{
The predicted bulge-to-total luminosity ratio in the rest-frame V-band 
for luminous red galaxies. The upper panel shows the predictions 
of the Baugh et~al. (dashed histogram) and Bower et~al. models (dotted 
histogram) at $z=0.24$. The lower panel displays the distributions 
predicted for LRGs at $z=0.50$. The distributions are normalized 
to give $\sum_i f_i = 1$.
}
\label{fig:bt}
\end{figure}

\begin{table}
\begin{center}
 \begin{tabular}{ccc}
 \hline
  Redshift & Baugh et~al. & Bower et~al. \\
 & [0,0.4]:[0.4,0.6]:[0.6,1.0] & [0,0.4]:[0.4,0.6]:[0.6,1.0] \\
 \hline
  $z = 0.24$ & 12:2:86 & 21:17:62 \\
  $z = 0.50$ & 21:6:73 & 24:22:54 \\
 \hline
 \end{tabular}
  \caption{
The predicted morphological mix of luminous red 
galaxies, at $z=0.24$ and $z=0.50$, for the Baugh et~al. 
and Bower et~al. models. The three values quoted for each model 
show the percentage of galaxies with bulge-to-total luminosity 
ratios of $B/T < 0.4$, $0.4 \le B/T \le 0.6$ and $B/T > 0.6$.
}
 \label{table:bt}
 \end{center}
\end{table}

\begin{figure}
{\epsfxsize=8.truecm
\epsfbox[18 144 592 718]{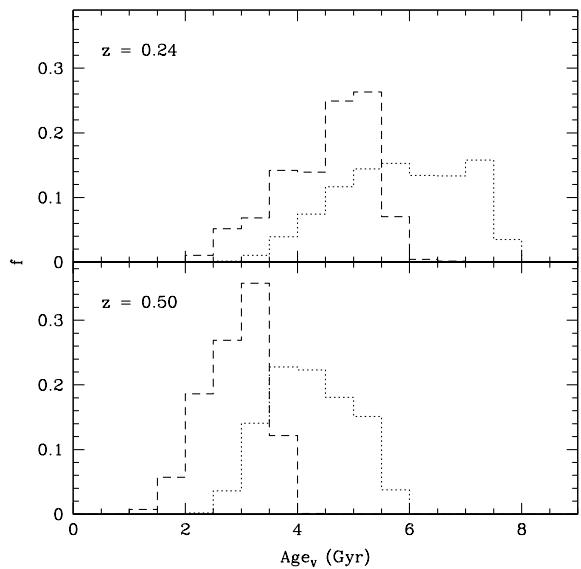}}
\caption{
The rest-frame V-band luminosity-weighted age of the stellar populations of 
LRGs at $z=0.24$ (upper panel) and $z=0.5$ (lower panel). As before, 
the predictions of the Baugh et~al. model are shown by the dashed 
histograms and those of the Bower et~al. model by the dotted histograms.
The histograms are normalized as in Fig.~\ref{fig:stellar.mass}.}
\label{fig:age}
\end{figure}

\begin{figure}
{\epsfxsize=8.truecm
\epsfbox[18 144 592 718]{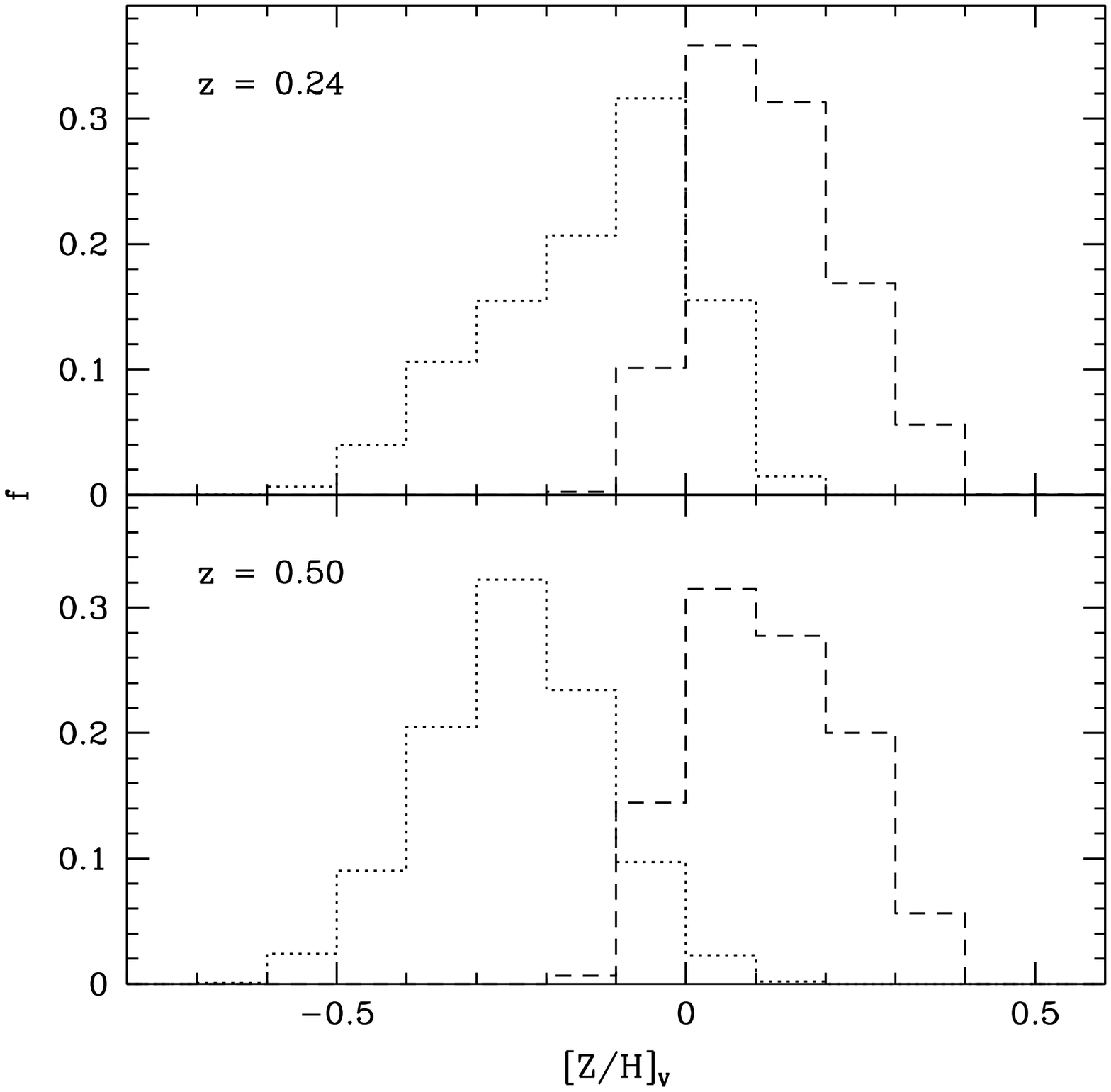}}
\caption{
The predicted distribution of the rest-frame V-band luminosity-weighted 
metallicity of LRGs. The upper panel displays the Baugh et~al. 
(dashed histogram) and Bower et~al. (dotted histogram) models 
at $z=0.24$ and the lower panel shows the predictions at $z=0.50$. 
The histograms are normalized as in Fig.~\ref{fig:stellar.mass}.
}
\label{fig:met}
\end{figure}

The bulge-to-total luminosity ratio, B/T, is often 
used as an indicator of the morphological type of a galaxy in 
semi-analytical models (see e.g. Baugh, Cole \& Frenk 1996). The B/T 
ratio is correlated with Hubble T-type, a subjective classification 
parameter relying upon the identification of features such as spiral 
arms and galactic bars, though there is considerable scatter around 
this relation (Simien \& de Vaucouleurs 1986). 
In the B-band, galaxies with $B/T <0.4$ correspond approximately to 
the T-types of late-type or spiral galaxies, those with $B/T > 0.6$ 
overlap most with elliptical galaxies and the intermediate values, 
$0.4 < B.T < 0.6$ correspond to lenticulars. 
In Fig.~\ref{fig:bt} we plot the predicted distribution of the 
bulge-to-total ratio in the rest-frame V-band, 
B/T$_{\rm V}$, for the Baugh et~al. and Bower et~al. models, at $z=0.24$ 
(upper panel) and $z=0.50$ (lower panel). The results are also summarized in 
Table~\ref{table:bt}, where the fraction of galaxies in intervals 
of B/T$_{\rm V}$ ratio are calculated.

In both models, the SDSS and 2SLAQ LRG samples are predicted to be 
mainly composed of bulge-dominated galaxies, which 
account for more than $\sim 60\%$ of the LRG population. However, 
the models suggest that the LRG samples contain an appreciable  
fraction ($\sim 20\%$) of late-type, disk-dominated systems. 
These galaxies meet the LRG colour selection criteria primarily 
because they have old stellar populations. Another prediction is 
that the fraction of bulge dominated galaxies is higher 
in the SDSS sample than in the 2SLAQ sample. 

The distributions of $B/T$ ratios predicted by the two models show 
substantial differences, particularly in the intermediate ratio range, 
which corresponds roughly to S0 types. This difference is not due to 
any single model ingredient, but is more likey to be the result of the 
interplay between several phenomena. As outlined in Section 2 (see also 
Almeida et~al. 2007), both models invoke galaxy mergers as a mechanism 
for making spheroids, either by the rearrangement of stellar disks or 
through triggering additional star formation. Bower et~al. also consider 
starbursts resulting from disks being dynamically unstable to bar formation.   
In the Baugh et~al. model, around 30\% of the total star formation takes 
place in merger driven starbursts. This figure is much lower in the 
Bower et~al. case, because galactic disks tend to be gas poor at high 
redshift in this model, as explained in Section 2. In the Bower et~al. 
model, starbursts resulting from the collapse of unstable disks dominate 
bursts driven by galaxy mergers. Baugh et~al. allow minor mergers 
to trigger starbursts, if the primary disk is gas rich. We have 
tested that removing these starbursts does not have a major impact 
on the distribution of $B/T$ values.

\subsubsection{Stellar populations}
\label{sec:popsyn}

\begin{figure*}
{\epsfxsize=14.truecm
\epsfbox[18 144 592 718]{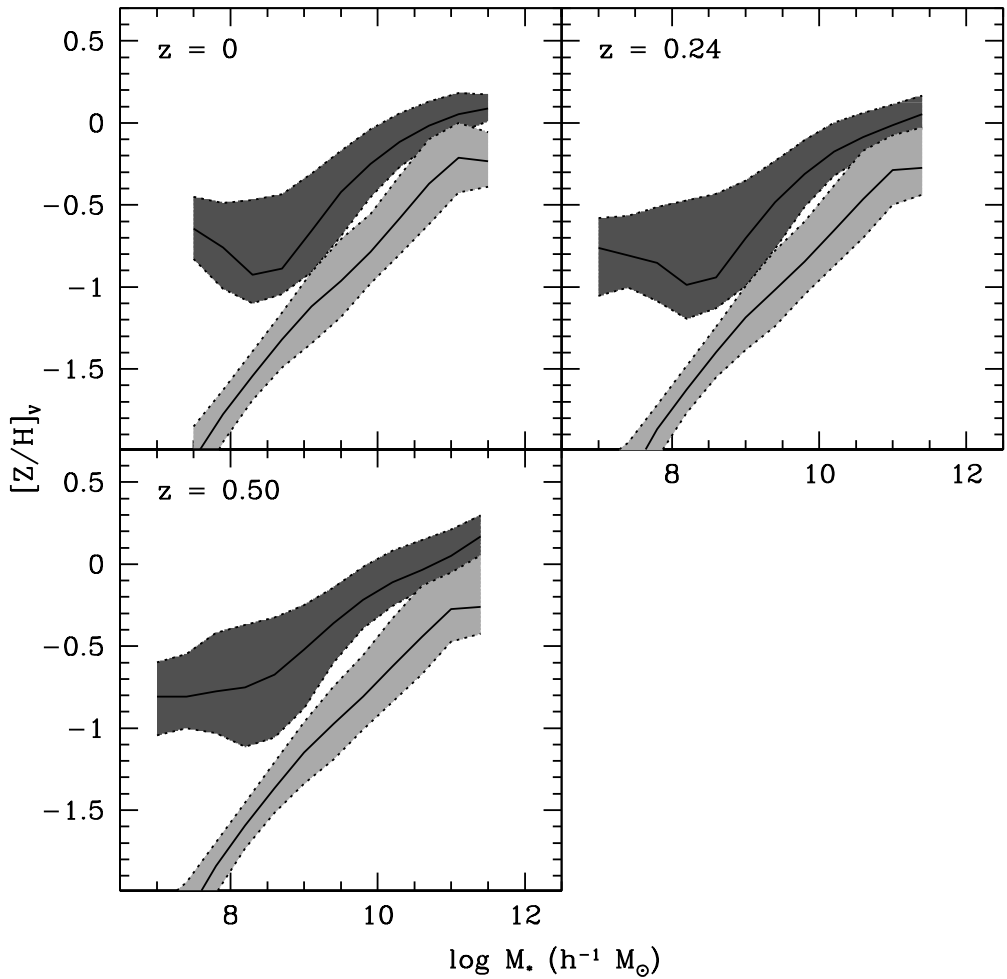}}
\caption{
The rest-frame V-band luminosity weighted metallicity -- stellar mass 
relation for ellipticals (i.e. galaxies with a bulge-to-total stellar 
mass ratio greater than 0.6). The light and dark gray shaded regions show 
the distributions for the Bower et al.~and the Baugh et al. models, 
respectively. Each panel corresponds to a different redshift, as indicated 
by the key. The solid line shows the median metallicity for 
stellar mass bins. The shaded regions are enclosed by the 10 
and 90 percentiles of the distribution.
}
\label{fig:metmstar}
\end{figure*}

The luminosity-weighted age of a stellar population is a measure  
of the age of the stars in a galaxy. The predicted distributions of the 
rest-frame V-band luminosity-weighted age of LRGs are plotted 
in Fig.~\ref{fig:age}. This plot reveals that LRGs have stellar 
populations with luminosity-weighted ages ranging from 4 to 8 Gyr 
in the Bower model, and from 2 to 6 Gyr in the Baugh et al. model, 
at $z=0.24$ (i.e. when the Universe was $\sim 80\%$ of its current age). 
At $z=0.50$ ($60\%$ of the current age of the Universe), 
galaxies in the Bower et~al. model show, again, older stellar ages than those 
in the Baugh et~al. model: the median of the distribution for LRGs 
in the Bower et~al. model is $\approx 4.2$ Gyr, whereas 
for the Baugh et~al. model it is $3.2$ Gyr. 
The difference in the age of the Universe between these redshifts 
is around $2$ Gyr (the value is slightly different for each model due to 
the different choice of the values of the cosmological parameters), and 
therefore accounts for the bulk of the difference in the ages of the 
SDSS and 2SLAQ samples.  
The model LRGs are therefore composed predominantly of old stellar 
populations and resemble those of observed early-type galaxies 
\citep[e.g.][]{trager, gallazzi}. Furthermore, our results 
are in excellent agreement with the analysis by \citet{barber}. 
Barber et~al. used stellar population 
synthesis models to fit the spectra of 4391 LRGs from the SDSS, 
finding matches for ages in the range from 2 to 10 Gyrs, with a peak in 
the distribution around 6 Gyrs. 
The difference in the ages predicted by the Baugh et~al. and 
Bower et~al. models has its origins in the different implementations of 
gas cooling and feedback applied in massive dark matter haloes. 
De Lucia et~al. (2006) show that the suppression of gas cooling due to 
AGN feedback tends to increase the age of the stellar population in 
the galaxies hosted by haloes with quasistatic hot gas atmospheres. 
Note that in hierarchical models, the age of the stars in a galaxy is not 
the same as the age of the galaxy: the age of the stellar population can 
greatly exceed the age of the galaxy, with stars forming in the galaxy's 
progenitors, which are later assembled into the final galaxy through mergers 
(e.g. Baugh et~al. 1996; Kauffmann 1996; De Lucia \& Blaizot 2007). We revisit 
this point in Section 5. 

Fig.~\ref{fig:met} shows the predicted distribution of the V-band 
luminosity-weighted metallicity for LRGs. 
There is little change in the luminosity weighted-metallicity between 
the two LRG samples in the Baugh et~al. model. In the Bower et~al. model, 
there is a modest decrease in metallicity of almost +0.2dex between 
$z=0.24$ and $z=0.5$. To help in the interpretation of these predictions 
it is instructive to plot the metallicity--stellar mass relation for 
spheroids at different redshifts (N.B. here, we consider any galaxy with 
a bulge-to-total stellar mass ratio in excess of 0.6, not just LRGs; we note 
that the metallicity--stellar mass relation is similar for galaxies with 
bulge-to-total ratios below 0.6). 
We recall that the typical stellar mass of LRGs is predicted to change 
by a factor of two between the $z=0.5$ and $z=0.24$ samples, 
from $\log(M_{*}/h^{-1}M_{\odot}) \sim 11$ to  $\log(M_{*}/h^{-1}M_{\odot}) 
\sim 11.3$.  
The predicted evolution of the stellar mass--metallicity 
relation for the Baugh et al.~and Bower et al.~models is shown in 
Fig.~\ref{fig:metmstar}. 
There is little evolution in the locus of the metallicity--mass 
relation. Between $z=0.5$ and $z=0.24$, the 
metallicity--stellar mass relation in the Baugh et~al. model 
flattens at the high mass end. Hence, the change in metallicity expected 
due to an increase in stellar mass by a factor of 2 using the 
metallicity--mass relation predicted at $z=0.5$ is largely cancelled 
out by the change in slope of the metallicity -- mass relation at $z=0.24$. 
In the Bower et~al. model, the metallicity--mass relation has a kink 
at $\log(M_{*}/h^{-1}M_{\odot}) \sim 11$, and is flat for the most 
massive galaxies over the whole of the redshift range plotted in 
Fig.~\ref{fig:metmstar}. 
The evolution seen in the Bower et~al. model is therefore due to the presence 
at $z=0.5$ of LRGs with masses $<10^{11}h^{-1}M_{\odot}$ which come from 
the steep part of the metallicity--mass relation; at $z=0.24$, only LRGs 
from the flat part of the metallicity--mass relation are sampled due to 
the increase in stellar mass.   

Fig.~\ref{fig:met} shows that the Bower et~al. model displays 
a different metallicity distribution from the Baugh et~al. model,  
predicting luminosity-weighted metallicities lower by a factor of $\sim 2$. 
This difference is entirely due to the choice of IMF used 
in the models. We remind the reader that, in the Baugh et~al. model, 
stars which form in merger driven bursts are assumed to be produced 
with a flat IMF, whereas in the Bower et~al. model, a \citet{ken} IMF is 
adopted in all modes of star formation.  The yield 
adopted is consistent with the choice of IMF. For a flat IMF, the yield 
is over six time larger than the yield expected from a Kennicutt IMF. 
As noted by \citet{nagashima}, the metal abundances for 
galaxies in the Baugh et~al. model are higher by a factor of 2-3 than 
is the case for a model using a Kennicutt IMF. Intriguingly, 
Barber et~al. (2007) also favour high metallicities in their simple fits to 
the spectra of SDSS LRGs, finding best fitting models in the 
range $-0.6 \la [{\rm Z/H}] \la 0.4$, which they argue is evidence 
in favour of LRGs forming stars with a top-heavy IMF.

\subsubsection{Are LRGs central or satellite galaxies?}
\label{sec:sat}

\begin{table}
\begin{center}
 \begin{tabular}{ccc}
 \hline
  Redshift & Baugh et~al. & Bower et~al. \\
 \hline
  $z = 0.24$ & 0.25 & 0.32 \\
  $z = 0.50$ & 0.21 & 0.30 \\
 \hline
 \end{tabular}
  \caption{
The predicted fraction of satellite galaxies in the 
luminous red galaxies samples at $z=0.24$ and $z=0.50$.
}
 \label{table:sat}
 \end{center}
\end{table}

In the models, the most massive galaxy within a dark matter halo is 
referred to as the ``central'' galaxy and any other galaxies which 
also reside in the halo are referred to as ``satellites'' (see 
Baugh 2006). In the majority of semi-analytical models, this distinction 
is important because gas which cools from the hot gas halo is directed 
onto the central galaxy, and satellite galaxies can merge only 
with the central galaxy. 
The fraction of luminous red galaxies which are 
satellite galaxies in the models is given in Table~\ref{table:sat}; 
more than 25\% of LRGs at $z=0.24$ are satellite galaxies 
with a slight decrease in this fraction for the 2SLAQ sample. 
The fraction of satellite galaxies has important consequences for the 
small scale clustering of LRGs, as we shall discuss in Section \S 4. 

\subsubsection{Radii}
\label{sec:radii}

\begin{figure}
{\epsfxsize=8.truecm
\epsfbox[18 144 592 718]{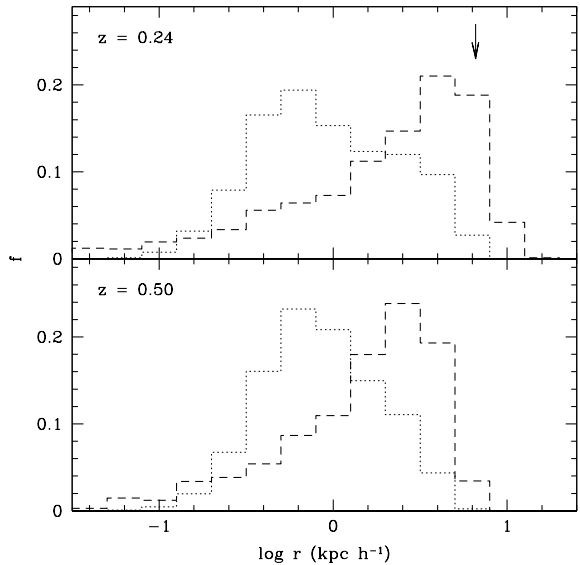}}
\caption{
The predicted half mass radii of luminous red galaxies. 
The upper and lower panels display the predictions of the Baugh 
et~al. (dashed histograms) and Bower et~al. models (dotted histograms), 
at $z=0.24$ and $z=0.5$ respectively. 
The histograms are normalized as in Fig.~\ref{fig:stellar.mass}. 
The arrow shows the median de Vaucouleur's radius of the observed 
SDSS LRGs.
}
\label{fig:radius}
\end{figure}


The distribution of the radii of model LRGs is shown in 
Fig.~\ref{fig:radius}. We plot the half-mass radius of the 
model galaxies, taking the mass weighted average of the 
disk and bulge components. The calculation of the linear sizes of 
galaxies in \g~takes into account the conservation of the angular 
momentum of cooling gas and the conservation of energy of 
merging galaxies (Cole et~al. 2000). 
This prescription  was tested against observations of bulge 
dominated SDSS galaxies by Almeida et~al. (2007). Overall, 
Almeida et~al. found that the predicted sizes of spheroids in the 
Baugh et~al. model matched the observed sizes reasonably well, except 
for galaxies much brighter than $L_{*}$. These bright galaxies 
are predicted to be a factor of up to three smaller than 
observed by \citet{bernardi_colour}. In the Bower et~al. model, the 
brightest spheroids are predicted to be even smaller than in the 
Baugh et~al. model. 
The same trend is seen in the predictions for the sizes of LRGs shown 
in Fig.~\ref{fig:radius}. At $z=0.24$, the median of the distribution of 
LRG half-light radii in the Baugh et~al. model occurs at 
$\sim 1.76 h^{-1}{\rm kpc}$. In the Bower et~al. model, this peak 
is at a radius that is around a factor of 2 smaller. At $z=0.5$, the 
median of the two model distributions differ by a smaller factor, 
$\approx 1.8$, although there is little evolution in the distributions 
from $z=0.5$ to $z=0.24$. 
The observed radii of SDSS LRGs (as extracted from the online database) 
are larger than the model predictions, 
with a median de Vaucouleur's radius of $6.6 h^{-1}{\rm kpc}$. 
The observational estimate of the LRG radius is obtained by fitting 
a de Vaucouleur's profile convolved with a seeing disk. In this case 
the seeing is restricted to be no worse than 1.4 arcseconds, which 
corresponds to a scale of $3.7 h^{-1}{\rm kpc}$ at the median redshift 
of the SDSS sample. Thus the tail of small scale-length galaxies predicted 
by the models would not be observable. However, the observed distribution 
of LRG sizes has few galaxies close to the seeing limit, so this is does 
not affect the estimation of the median size of SDSS LRGs. 

\subsubsection{Velocity dispersion} 
\label{sec:sigma}

\begin{figure}
{\epsfxsize=8.truecm
\epsfbox[18 144 592 718]{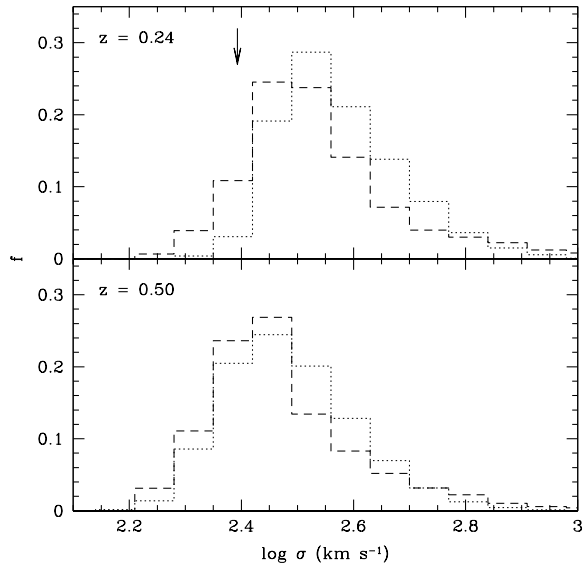}}
\caption{
The predicted one-dimensional velocity dispersion of 
luminous red galaxies in the Baugh et~al. (dashed lines) 
and Bower et~al. models (dotted lines). The upper panel shows  
the distributions at $z=0.24$ and the lower panel at $z=0.50$. 
The histograms are normalized as in Fig.~\ref{fig:stellar.mass}. 
The median velocity dispersion of the observed SDSS LRGs is 
represented by the arrow.
}
\label{fig:sigma}
\end{figure}

Fig.~\ref{fig:sigma} shows the distribution of the one-dimensional 
velocity dispersion of the bulge component of model LRGs, $\sigma_{\rm 1D}$. 
This is calculated from the effective circular velocity of the bulge, 
$V_{\rm c, bulge}$, using $\sigma_{\rm 1D} = (1.1/\sqrt{3})V_{\rm c, bulge}$, 
where $\sigma_{\rm 1D}$ is assumed to be isotropic. 
The circular velocity at the half mass radius of the bulge is a model 
output which is computed taking into account the angular momentum and 
mass of the disk and bulge, and the gravitational contribution of the 
baryons and dark matter (see Cole et~al. 2000 for further details).   
The factor of $1.1$ is an empirical adjustment introduced by 
Cole et~al. (1994) which we have retained to facilitate comparison 
with predictions for the more general population of spheroids 
presented in Almeida et~al. (2007). 
At $z=0.24$, both models predict velocity dispersions in the range 
$220$--$400{\rm km s}^{-1}$, with a median around $320 {\rm km s}^{-1}$.  
Between $z=0.24$ and $z=0.5$, the predicted distribution of LRG 
velocity dispersions shifts to lower values by $\Delta \log \sigma 
\approx 0.1$. 
The bulk of this evolution is due to the change in stellar mass between 
the LRG samples \citep[see Fig. 16 of][]{almeida}.
The median velocity dispersion for SDSS LRGs is 
$\sigma = 250 {\rm km s}^{-1}$, which is somewhat smaller than 
the model predictions. 

\subsection{Why do the two models give different predictions?} 
\label{sec:diff }

In \S3.2, we demonstrated that the predictions of the Bower et~al. model 
for the luminosity function of LRGs are in better agreement with the 
observations than those of the Baugh et~al. model. In particular, 
at $z=0.24$, the Bower et~al. model gives a very good match both to 
the shape of the observed luminosity function and the integrated number 
density of LRGs, matching the observed density at the 10\% level. 
The Baugh et~al. model predicts too many LRGs at this redshift, 
particularly at the bright end. 

We saw in Section 2 that there are several areas in which the input 
physics and parameter choices differ between the two models. Although 
our aim in this paper is to test published models and not to tweak the 
results to fit the LRG population, it is instructive to vary some of 
the parameters in the Baugh et~al. model to see if the predictions 
for the number of LRGs improve. We varied several model 
ingredients (e.g. strength of superwind feedback, star formation 
timescale, burst duration, choice of IMF in bursts, criteria for 
triggering a starburst following a galaxy merger) and found that in 
all cases, the resulting change in the luminosity function of LRGs 
was driven by a change in the overall luminosity function, i.e. if the 
number of bright LRGs increased, the luminosity function of all galaxies 
was found to brighten by a similar amount. Since a model is only deemed 
successful if it reproduces as closely as possible the overall galaxy 
luminosity function, none of these variant models is acceptable without 
further parameter changes to reconcile the overall luminosity function 
with observations. Hence, the apparent gain in the abundance of 
LRGs will be cancelled out by the additional parameter changes which 
compensate for the brightening of the overall luminosity function. 
We note that adopting a hot gas density profile with a fixed rather 
than evolving core radius in the Baugh et~al. model does not improve 
the predictions for the abundance of LRGs. With a fixed core radius, 
more gas cools in massive haloes than in the evolving core case, which 
leads to more bright galaxies (see Cole et al. 2000). However, these 
galaxies are also bluer and so do not match the LRG selection. 

The success of the Bower et~al. model can be traced to the revised 
gas cooling prescription adopted in massive haloes. The suppression 
of the cooling flow in haloes with a quasistatic hot atmosphere and the 
dependence of this phenomenon on redshift are the key reasons why this 
model matches the evolution of the LRG luminosity function better than 
the Baugh et~al. model. LRGs in the Bower et~al. model are older than 
their counterparts in the Baugh et~al. model, because the supply of cold 
gas for star formation is removed. In the Baugh et~al. model, the superwind 
feedback acts to effectively suppress star formation in massive haloes, 
but still allows some star formation to take place. The choice of a 
top-heavy IMF in starbursts helps the Baugh et~al. model, to mask the  
recent star formation to some extent, by making the LRG stellar 
population more metal rich and thus redder.

\section{The clustering of LRGs}

The clustering of galaxies is an invaluable constraint on theoretical 
models of galaxy formation. The form and amplitude of the two-point 
galaxy correlation function is driven by three main factors, which play 
different roles on different length scales: the clustering of the 
underlying dark matter, the distribution or partitioning of galaxies 
between dark matter haloes (e.g. Benson et~al. 2000; Seljak 2000; 
Peacock \& Smith 2000) and the distribution of galaxies within haloes. 
The number of galaxies as a function of halo mass, 
called the halo occupation distribution (HOD), controls how the correlation 
function of galaxies is related to that of the underlying matter 
(see the review by Cooray \& Sheth 2002). 
On large scales, the correlation functions of the galaxies and matter 
have similar shapes, but differ in amplitude by the square of the effective bias. 
On smaller scales, comparable to the radii of the typical 
haloes which host the galaxies of interest, it is the number of galaxies 
and its radial distribution within the same dark matter halo which sets 
the form and amplitude of the correlation function (see Fig.10 of Benson et~al. 2000). 
The clustering of the galaxies can be different from that of the matter 
on small scales as well as large. 
The predictions for the correlation function in redshift-space 
can also be affected by the peculiar motions of galaxies as we 
shall see later on in this section.

\begin{figure}
{\epsfxsize=8.truecm
\epsfbox[18 144 592 718]{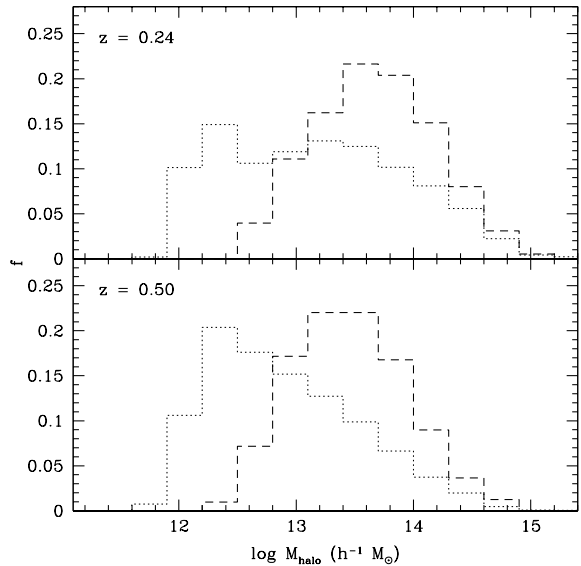}}
\caption{
The distribution of the masses of haloes which host LRGs at  
$z=0.24$ (upper panel) and $z=0.5$ (lower panel). As before, 
the dashed histogram shows the predictions of the Baugh et~al. 
model and the dotted line shows the Bower et~al. model. 
The histograms are normalized as in Fig.~\ref{fig:stellar.mass}.}
\label{fig:halo.mass}
\end{figure}

Semi-analytical models naturally predict which dark matter haloes 
contain LRGs. Fig.~\ref{fig:halo.mass} shows the range of dark halo 
masses that host LRGs in the Baugh et~al. and Bower et~al. models. 
Far from being restricted to cluster-mass haloes, in the models LRGs   
can occur in a wide range of halo masses, including fairly modest haloes 
comparable in mass to the halo which is thought to host the Milky Way. 
This plot reveals important differences between the predictions of 
the two models. At $z=0.24$, LRGs in the Bower et~al. model occupy haloes 
with masses in the range $1\times10^{12} - 1\times10^{15}h^{-1}M_{\odot}$, 
with a median of $\sim 1\times10^{13}h^{-1}M_{\odot}$. 
The Baugh et~al. model predicts that LRGs are to be found in haloes which 
are a factor of $\sim 3$ more massive than in the Bower et~al. model, 
with the median of the distribution occuring at 
$3\times10^{13}h^{-1}M_{\odot}$. The prediction from the Baugh et~al. 
model is in excellent agreement with the halo mass estimated for 
SDSS LRGs using weak lensing measurements (Mandelbaum et~al. 2006).  
Despite the differences in the sample selection, there is little 
evolution with redshift in the distribution of host halo masses 
from $z=0.24$ to $z=0.5$, with a slight shift to lower halo masses seen 
at the higher redshift. The difference in the range of halo 
masses predicted to host LRGs will have an impact on the amplitude of the LRG 
correlation function, and, thus a measurement of the clustering of LRGs 
can potentially discriminate between the two models. 

\begin{figure*}
{\epsfxsize=14.truecm
\epsfbox[18 144 592 718]{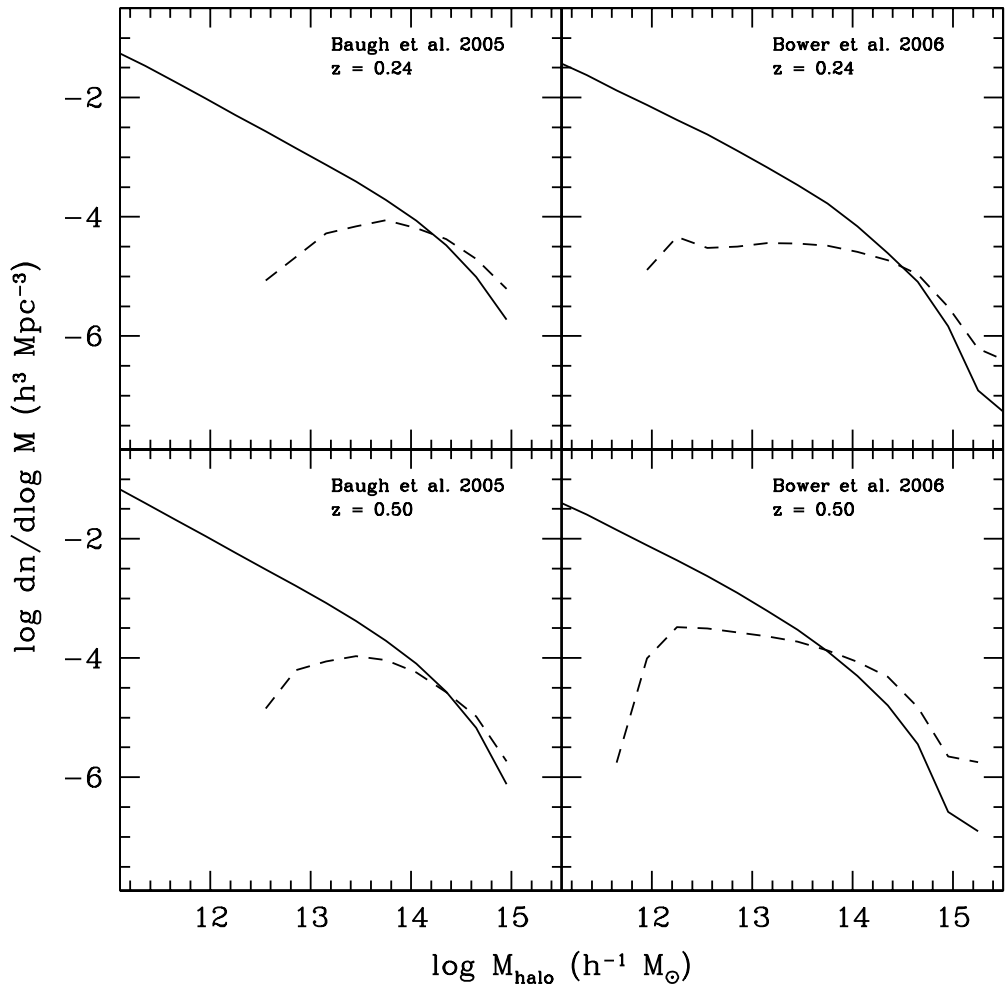}}
\caption{
The mass function of dark haloes (solid lines) and of haloes which 
host an LRG (dashed lines). The contribution of each halo to the 
mass function plotted with a dashed line is directly proportional 
to the number of LRGs it contains; where the dashed line has a greater 
amplitude than the solid line, this means that haloes of that mass contain 
more than one LRG on average. The left-hand panels show the 
predictions for the Baugh et~al. model and the right-hand panels 
for the Bower et~al. model. The top row corresponds to $z=0.24$ and 
the bottom row to $z=0.5$.
}
\label{fig:mass.function}
\end{figure*}

\begin{figure*}
{\epsfxsize=14.truecm
\epsfbox[18 144 592 718]{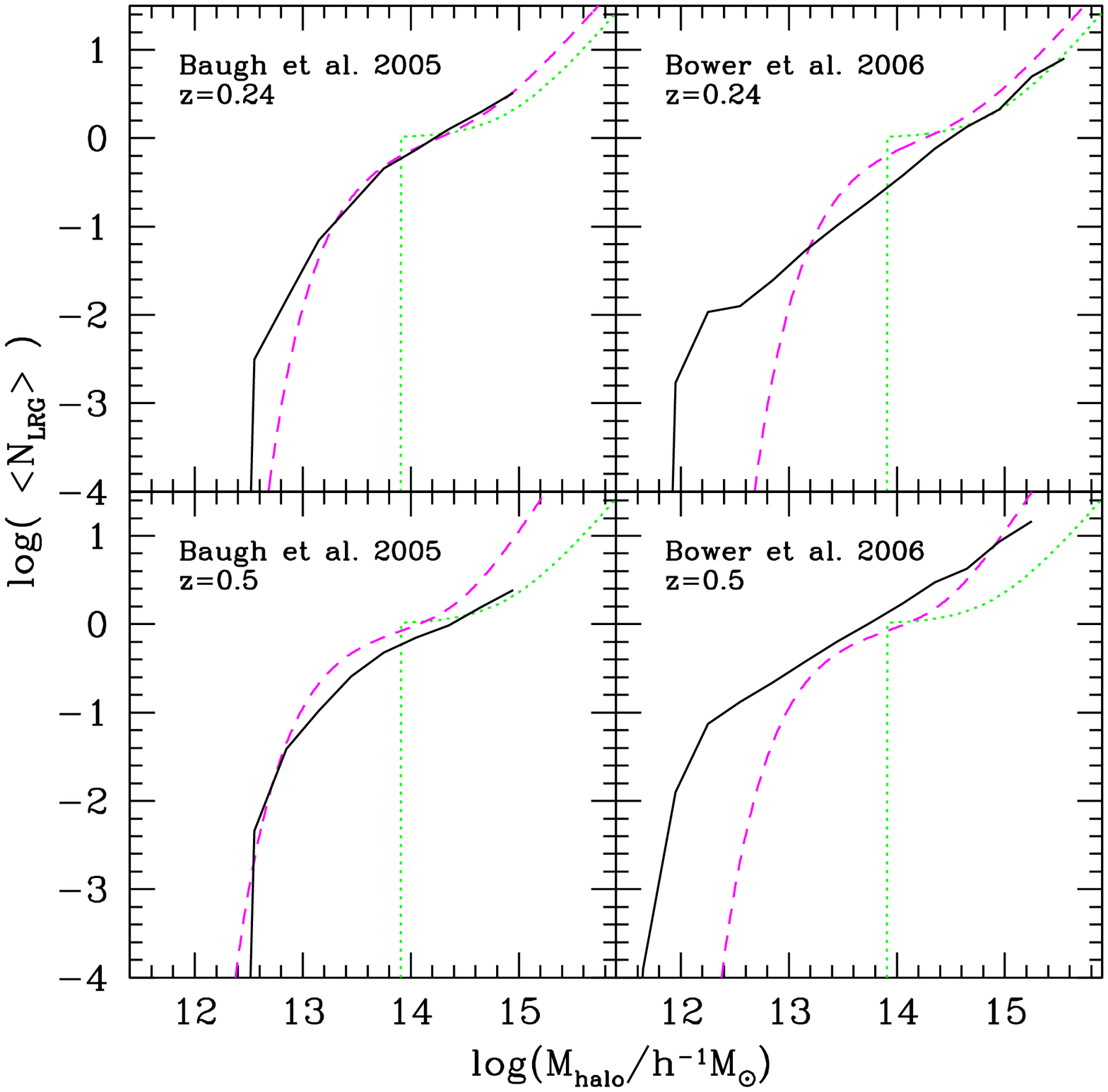}}
\caption{
The halo occupation distributions (HOD) of LRGs predicted by the models.  
The left-hand panels show the predictions for the Baugh et~al. 
model and the right-hand panels for the Bower et~al. model. The 
top row corresponds to $z=0.24$ (SDSS) and the bottom row to $z=0.5$ (2SLAQ).
The dotted line shows the HOD fit used by Masjedi et~al. to describe 
SDSS LRGs, and is reproduced in each panel.
The dashed lines show the HOD fit to the SDSS and 2SLAQ samples 
obtained by Wake et~al. (2008).
}
\label{fig:hod}
\end{figure*}

The distribution of halo masses hosting LRGs plotted 
in Fig.~\ref{fig:halo.mass} is determined by two factors: 
the adundance of dark matter haloes, which is a 
strong function of halo mass for the typical hosts of LRGs, and the 
number of LRGs which occupy the same dark halo. 
These factors are separated in Fig.~\ref{fig:mass.function}, in which 
we compare the overall mass function of dark haloes with the mass function of 
haloes weighted by the number of LRGs they contain. 
The solid lines show the mass function of dark matter haloes, which is 
determined by the values of the cosmological parameters, the form and 
amplitude of the power spectrum of matter fluctuations and the redshift 
(e.g. Governato et~al. 1999). 
The dashed lines show the mass function of haloes multiplied by the average 
number of LRGs as a function of halo mass.  
At the mass where the solid and dashed lines cross, these haloes host on 
average one LRG. At lower masses, the mean number of LRGs per halo rapidly 
falls below one (this is the ratio of the abundance indicated by the 
dashed line divided by the abundance shown by the solid line). There is a
threshold mass which must be reached before there is any possibility of 
a halo hosting an LRG. At this mass, only a tiny fraction of haloes actually 
contain an LRG. Nevertheless, these low mass haloes, because they are 
much more abundant than the more massive haloes, which have a higher mean 
number of LRGS, make an important contribution to the overall space 
density of LRGs. Hence it is essential for a model to take into account the 
scatter in the formation histories of dark matter haloes in order to 
accurately model the space density and clustering of LRGs. In the high mass 
tail of the mass function, the amplitude of the dashed curve exceeds that 
of the solid curve; at these masses, haloes host more than one LRG. 
Fig.~\ref{fig:mass.function} shows that, at $z=0.24$, in 
the Baugh et~al. model there are an average of $\sim 3$ SDSS LRGs per 
halo at ${\rm M_{halo}} \approx 1\times 10^{15}\,{\rm h^{-1} M_{\odot}}$. 
At $z=0.5$, the mass of haloes which host on average one LRG 
(${\rm M_{halo}} \approx 2.5\times 10^{14}\, {\rm h^{-1} M_{\odot}}$) 
is higher than at $z=0.24$ and the most massive haloes do not contain 
as many LRGs as the most massive haloes present at $z=0.24$. 
This figure also shows that the Bower et~al. model predicts 
a higher mean number of LRGs per halo than the Baugh et~al. 
model at $z=0.5$; for haloes of mass ${\rm M_{halo}} \approx 1\times 10^{15}\, 
{\rm h^{-1} M_{\odot}}$, the Bower et~al. model predicts an average 
of 10 LRGs per halo. The multiple occupancy of LRGs in high mass haloes 
has important consequences for the form of the predicted 
correlation function at small pair separations (i.e. $r < 1 h^{-1}$Mpc).   

\begin{figure}
{\epsfxsize=8.truecm
\epsfbox[18 144 592 718]{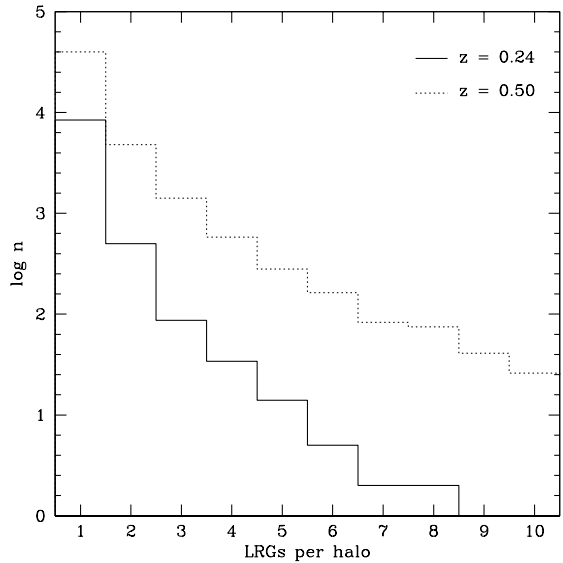}}
\caption{
The number of times a given number of LRGs is found 
within a common host halo in the Bower et~al. model, 
in the whole of the Millennium Simulation volume. 
(Note we do not show haloes with zero LRGs.)  
The solid and dotted lines show the distributions 
at $z=0.24$ and $z=0.50$, respectively. Recall that the space density of 
LRGs selected at $z=0.5$ is greater than that at $z=0.24$. Note 
that the distribution of occupation numbers at $z=0.50$ 
extends out to around 30 LRGs per halo.
}
\label{fig:nlrgs}
\end{figure}

\begin{figure}
{\epsfxsize=8.truecm
\epsfbox[18 144 592 718]{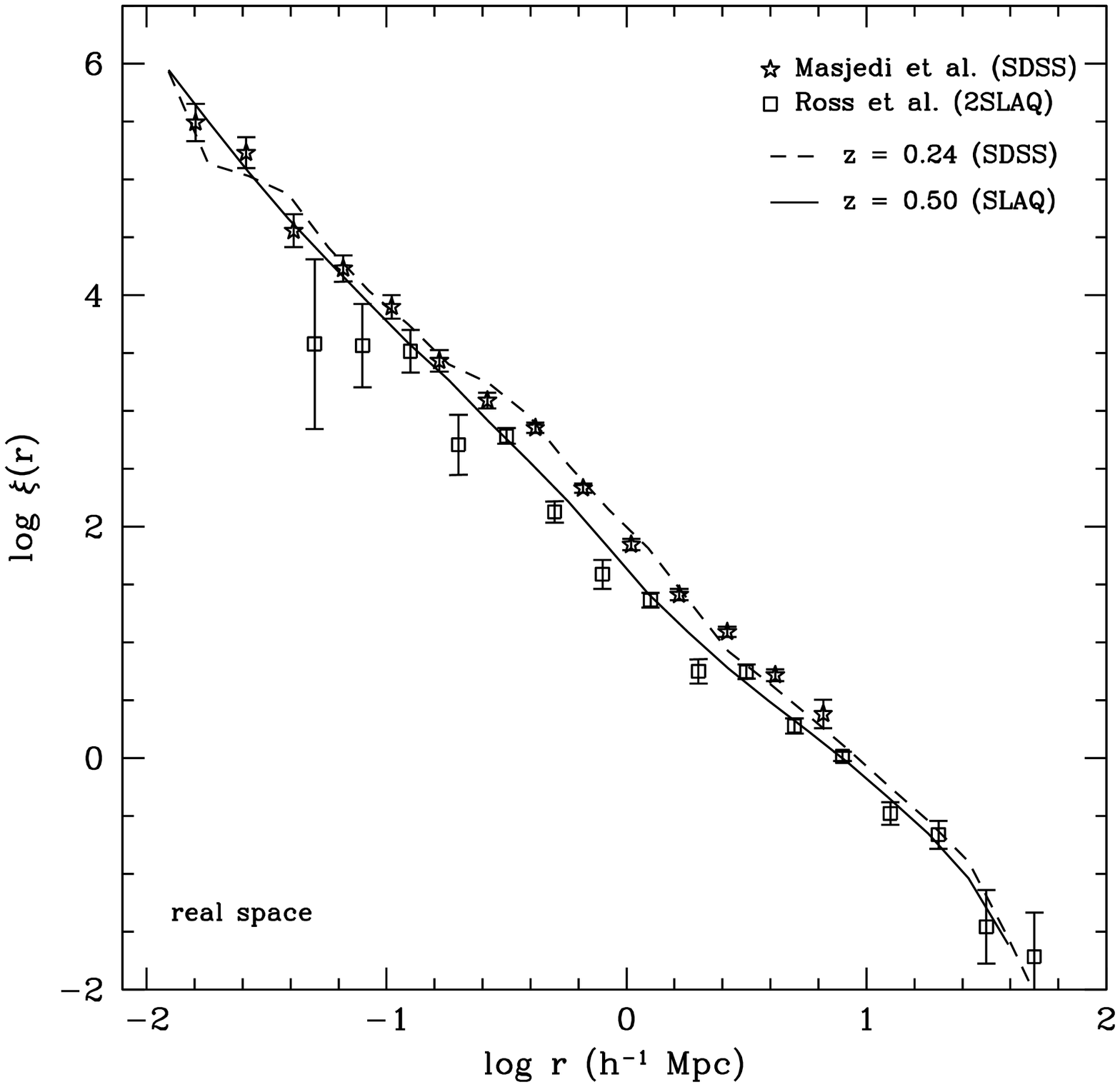}}
{\epsfxsize=8.truecm
\epsfbox[18 144 592 718]{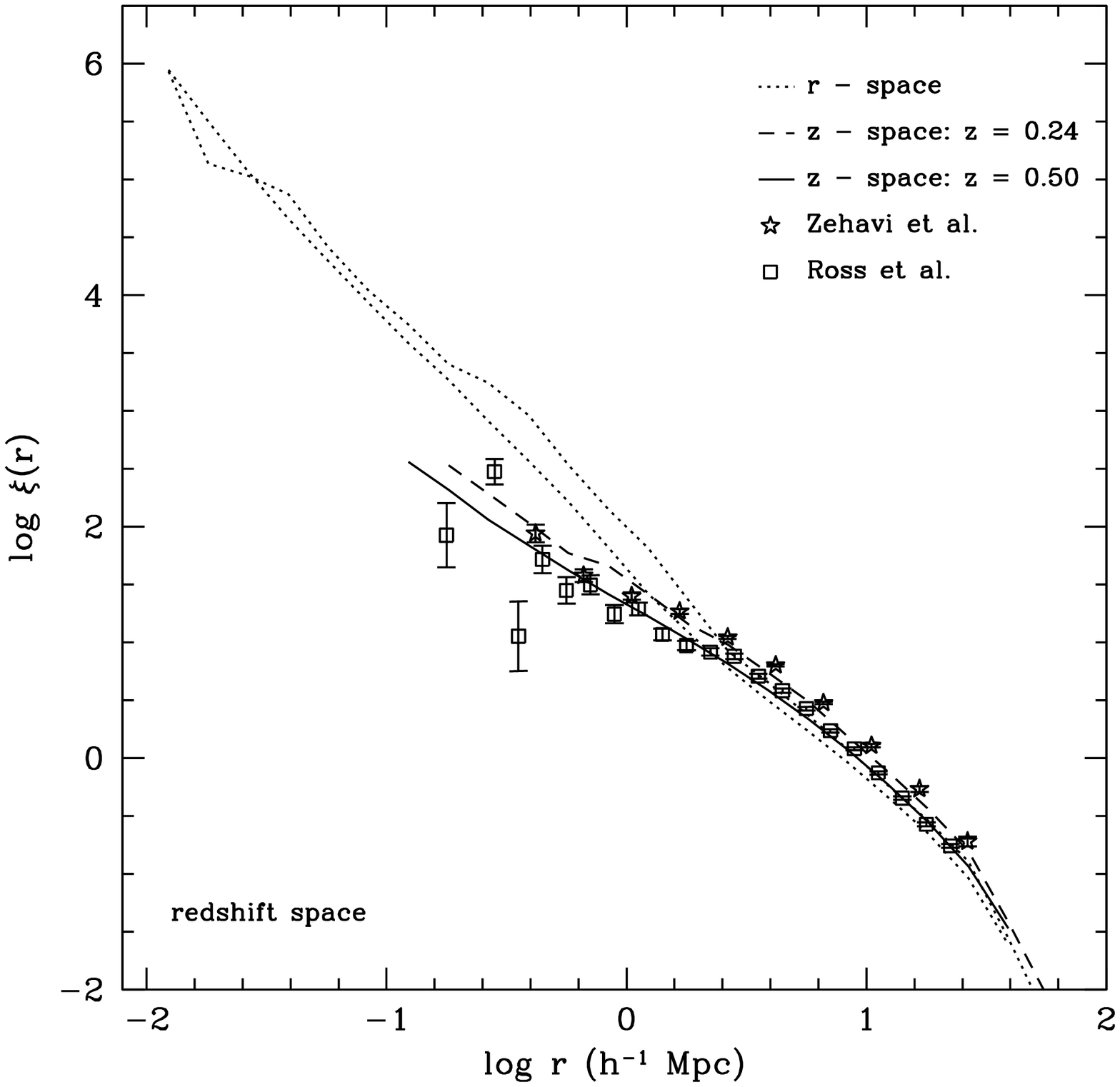}}
\caption{
The two-point correlation function of LRGs in the 
Bower et~al. model.
{\it Upper panel:} The real-space correlation function at $z=0.24$ 
(dashed line) and $z=0.50$ (solid line). Also shown in this panel 
are data from \citet{masjedi} (SDSS LRG sample) and from \citet{ross} 
(2SLAQ sample). 
{\it Lower panel:} The redshift-space correlation function, at $z=0.24$ 
(dashed line) and $z=0.50$ (solid line). For comparison we plot the 
real-space correlation functions shown in the upper panel using dotted 
lines. The symbols show the correlation function of LRGs 
in redshift-space from \citet{zehavi} and \citet{ross}.
}
\label{fig:xi}
\end{figure}

\begin{figure}
{\epsfxsize=8.truecm
\epsfbox[18 144 592 718]{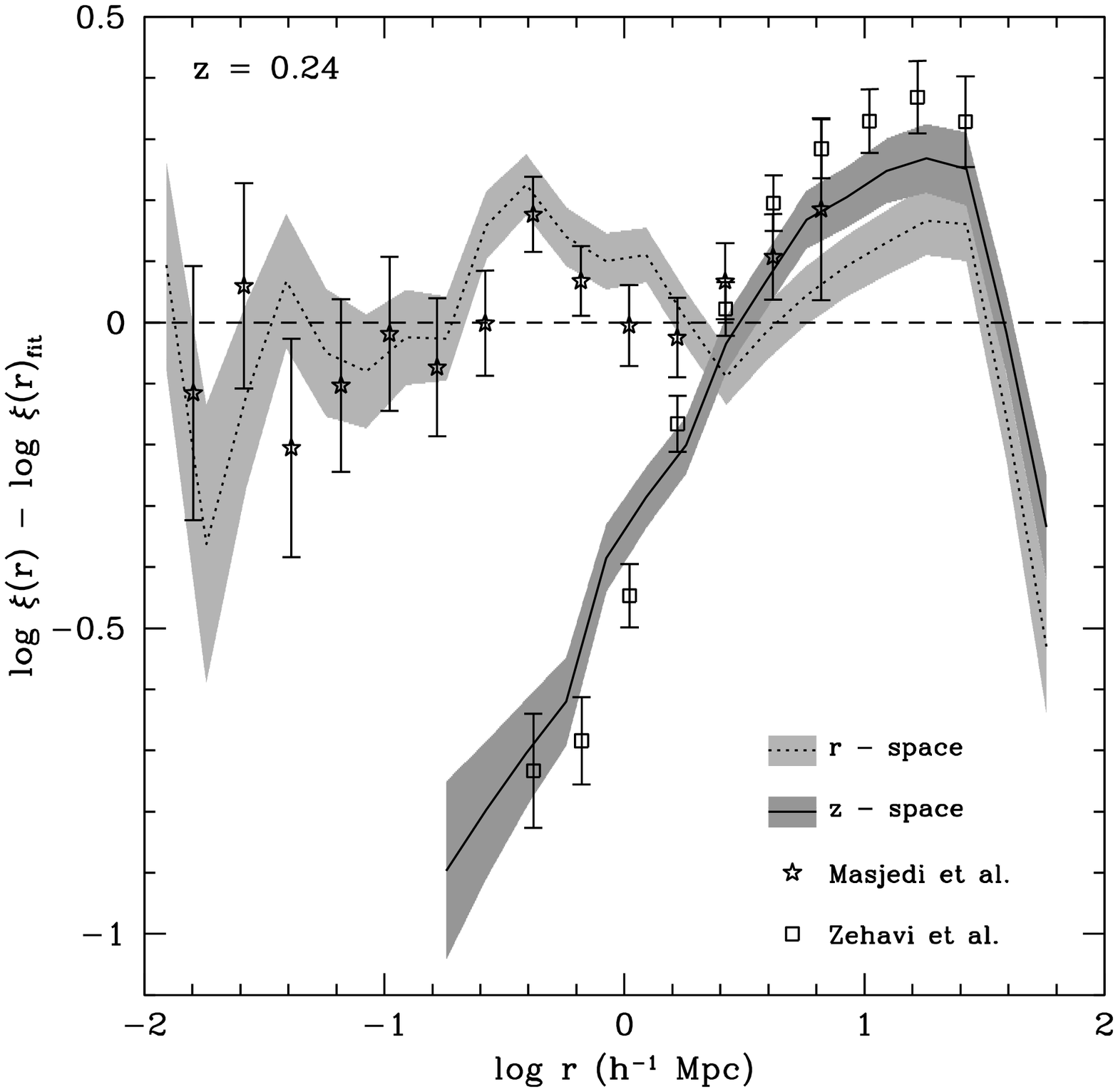}}
{\epsfxsize=8.truecm
\epsfbox[18 144 592 718]{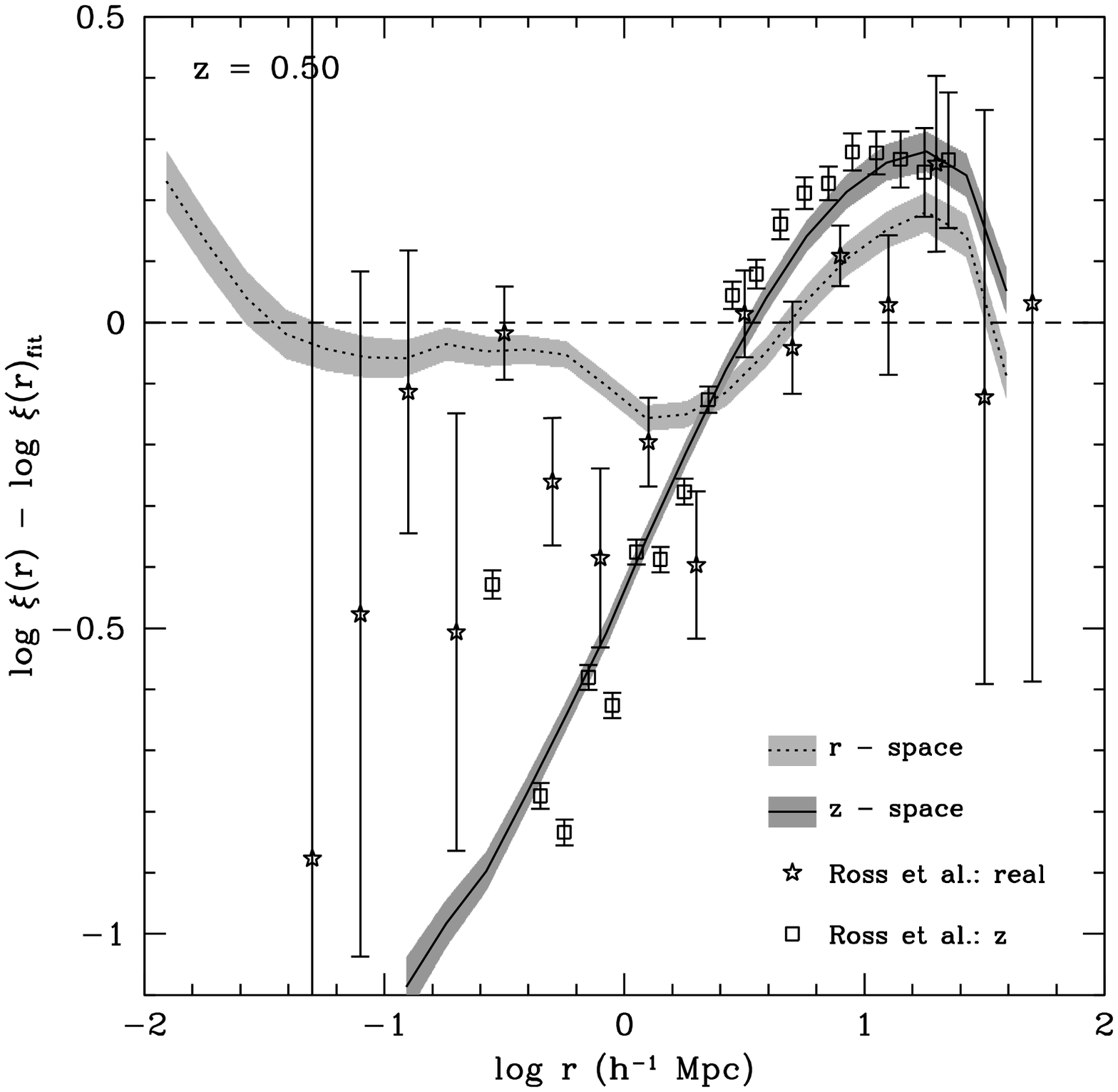}}
\caption{
The ratio between the two-point correlation function of LRGs and a 
power law, $\xi(r)_{\rm fit} = (r/r_{0})^{-2.07}$, 
where $r_0 = 8.2\,h^{-1}\,{\rm Mpc}$ at $z=0.24$ (upper panel) 
and $r_0 = 7.1\,h^{-1}\,{\rm Mpc}$ at $z=0.50$ (lower panel). 
The dotted and solid lines show the predictions of the Bower et~al. 
model in real and redshift-space, respectively. 
The shaded areas show the $1\sigma$ Poisson errors derived using the 
number of pairs expected in the model at a given separation.
}
\label{fig:xifit}
\end{figure}

The mean number of LRGs as a function of halo mass in the halo occupation 
distribution (HOD), as predicted by \g, is plotted in Fig.~\ref{fig:hod}. For 
comparison, we also plot the function quoted as a description of the HOD for 
SDSS LRGs by Masjedi et~al. (2006), which is reproduced in each panel of 
Fig.~\ref{fig:hod} to serve as a reference point (dotted line). 
The Masjedi et al.~HOD was not derived by fitting the model correlation 
functions to that measured for LRGs. Instead, this is simply the 
HOD for the brightest luminosity bin of the main SDSS galaxy sample 
analyzed by Zehavi 
et~al. (2005). Although galaxies in this sample have similar luminosities 
and colours to LRGs, they have a different redshift distribution and the 
selection is only crudely matched. Fig.~\ref{fig:hod} shows that whilst this 
parametric form for the HOD is a reasonable match to that predicted 
by the models for massive haloes with more than 1 LRG, it is a poor 
description at lower masses. As we have argued before, even though 
low mass haloes host a mean number of LRGs below unity, there are more 
of them than there are high mass haloes, so these objects make a 
significant contribution to the clustering signal (see a similar discussion 
in Baugh et~al. 1999). 
Ho et~al. (2008) estimated the HOD for a sample of 
LRGs in clusters and found a result similar to the predictions 
of the Bower et~al. model at $z=0.5$.  
A more detailed modelling of the transition from 
1 LRG per halo to 0 LRGs per halo is required to describe the model 
predictions, such as that advocated by Zheng et~al. (2005). 
Wake et~al. (2008) have carried out such a calculation, fixing the 
background cosmology to match that used in the two models, and 
Fig.~\ref{fig:hod} shows that their estimates are in better agreement 
with the model predictions (Baugh et al.~model) than with the HOD advocated 
by Masjedi et~al. (2006; see also Kulkarni et~al. 2007). 
Nevertheless, there are still some discrepancies between the fit obtained 
by Wake et~al. and our model predictions. This could be due to the fact 
that the models do not reproduce exactly the number density of LRGs, 
whereas Wake et~al. include this as a constraint on their HOD parameters.

We plot the frequency of finding a given number of LRGs within a 
common dark matter halo in Fig.~\ref{fig:nlrgs}. Here the number 
of haloes plotted on the y-axis is the number within the full volume 
of the Millennium simulation ($0.125h^{-3}{\rm Gpc}^{3}$), 
which contain the specified number of LRGs. At $z=0.5$, nearly ten 
thousand dark matter haloes in the Millennium contain more 
than one LRG. Haloes with only 
one LRG are approximately ten times more common. At the median 
redshift of the 2SLAQ survey, the tail of haloes with more than 
one LRG extends to $\sim 30$, reflecting the higher 
space density of LRGs in this sample compared with the SDSS LRG sample.

Before presenting explicit predictions for the correlation function 
of LRGs, it is instructive to compute the asymptotic bias factor, 
$b_{\rm eff}$, which quantifies the boost in the clustering of LRGs 
relative to that of the underlying matter distribution 
on large scales ($\xi_{\rm LRG} \approx b^{2}_{\rm eff} \xi_{\rm mass}$). 
This will allow us 
to compare the clustering predictions of the Baugh et~al. and 
Bower et~al. models. (We cannot make a direct prediction of the 
correlation function of galaxies in the Baugh et~al. model, because, 
unlike the Bower et~al. model, it is not implanted in an N-body 
simulation.) The asymptotic bias factor can be calculated analytically 
using the mass function of haloes which host an LRG, $N(z, M')$,  
and the bias as a function of halo mass, $b(M',z)$ (e.g. Baugh et~al. 1999): 
\begin{equation}
 b_{\rm eff}(z) = \frac{\int_{M} N(z, M')\,b(M',z)\,{\rm d}\ln M'}{
\int_{M} N(z, M'){\rm d}\ln M'}.
\end{equation}
The integrals are over the range of halo masses which host LRGs. 
The bias factor, $b(M, z)$, for haloes of mass $M$, as a function of redshift is 
computed using the prescription of Sheth, Mo \& Tormen (2001). For the Baugh 
et~al. model we calculate an effective bias of $b_{\rm eff}=2.45$ at $z=0.24$, and 
$b_{\rm eff}=2.27$ at $z=0.50$. In the case of the Bower et~al. model the values 
are slightly lower, with $b_{\rm eff} = 1.82$ at $z=0.24$ and 
$b_{\rm eff} = 1.72$ at $z=0.50$. Using a sample of 35\,000 LRGs from 
the SDSS, Zehavi et al. (2005) measured a bias of $b=1.84\pm 0.11$, 
which is in agreement with the predictions of the Bower et~al. 
model (see also Kulkarni et al. 2007 and Blake et al. 2007).

For the remainder of this section, we focus on the predictions of the 
Bower et~al. model. As this model is implemented in an N-body simulation, 
it can be used to produce direct predictions for the spatial distribution 
of galaxies and hence the two-point correlation function. The published 
Bower et~al. model associates the central (biggest) galaxy in each halo with the 
largest substructure in the halo. For haloes without resolved substructures,
the central galaxy is assigned to the position of the most bound particle. 
Satellite galaxies are associated with the substructure corresponding to 
the halo in which they formed or to the most bound particle from the halo
in which they formed. Fig.~\ref{fig:xi} shows the predicted 
correlation function for this model in real-space and in redshift-space. In real-space, 
the cartesian co-ordinates of the LRGs within the simulation box are used 
to compute pair separations. In the redshift-space, 
galaxy positions along one of the axes are perturbed by the peculiar 
velocity of the galaxy, scaled by the appropriate value of the Hubble 
parameter. This corresponds to the distant observer approximation, 
which is reasonable given the median redshifts of the observational samples. 
Fig.~\ref{fig:xi} shows that the correlation functions predicted by the 
Bower et~al. model agree spectacularly well with the measured correlation 
functions, both in real and redshift space. The agreement between the model 
predictions and the observational estimates in real-space is 
particularly noteworthy. The real-space correlation function of LRGs 
is very close to a power law over three and a half decades in pair 
separation, varying in amplitude over this range by nearly eight orders 
of magnitude. The extension of the power-law in the model predictions 
from $r \sim 1h^{-1}$Mpc down 
to $r \approx 0.01 h^{-1}$Mpc is a remarkable success of the model. 
The correlation function on these scales is determined by pairs of LRGs 
within the same dark matter halo. If the model did not predict that some 
haloes contain more than one LRG, the correlation function would tend 
to $\xi \sim -1$ on scales smaller than the typical radius of the haloes 
hosting LRGs (see Benson et~al. 2000 for a discussion of this point). The 
slope of the correlation function on such small scales is a strong test 
of the model through the predicted number of LRGs per halo. 

In the lower panel of Fig.~\ref{fig:xi}, we have retained the same dynamic 
range on both axes to allow a ready comparison of the clustering signal 
predicted in redshift-space with that obtained in real-space. To further aid 
this comparison, we have also reproduced the real-space predictions from 
the upper panel as dotted lines. The impact of including the contribution of 
peculiar motions when inferring the distance to galaxies depends on the 
scale. On intermediate and larger scales($r > 3 h^{-1}$Mpc), bulk motions 
of galaxies result in an enhancement in the amplitude of 
the correlation function 
measured in redshift-space. This boost is modest because, as we demonstrated 
above, LRGs are biased tracers of the matter distribution (Kaiser 1987). 
On small scales, the clustering signal in redshift-space is significantly 
lower than in real-space. Again, this feature of the predictions, a damping 
of the clustering on small scales in redshift-space, is expected if 
the sample contains haloes which host multiple LRGs; the peculiar motions 
of the LRGs within the halo cause an apparent stretching of the structure 
in redshift-space, diluting the number of LRG pairs. 
The clustering predicted in redshift-space 
agrees extremely well with the measurements by Zehavi et~al. (2005) and Ross 
et~al. (2007a). 

Another view of the comparison between the predicted and measured correlation 
functions is presented in Fig.~\ref{fig:xifit}, in which we plot the 
correlation function divided by a reference power law,  
$\xi(r)_{\rm fit} = \left(r/r_{0}\right)^{\gamma}$.  
This way of plotting the results emphasizes 
any differences between model and data by expanding the useful dynamic  
range plotted on the y-axis. In both panels, for the reference power law 
we fit a slope of  $\gamma = -2.07$, which agrees with the slope 
inferred for the real-space correlation function by Masjedi et~al. (2006). 
For the correlation lengths in each panel, we use   
$r_0 = 8.2\,{\rm h^{-1}\,Mpc}$ at $z=0.24$ (upper panel) and 
$r_0 = 7.1\,{\rm h^{-1}\,Mpc}$ at $z=0.50$ (lower panel). 
Fig.~\ref{fig:xifit} shows clearly the difference between the shape of 
the correlation function in real-space and redshift space. The level of 
agreement between the model predictions and the measurements is impressive, 
particularly in view of the fact that no model parameters were 
fine-tuned to achieve this match. 

\section{The star formation and merger histories of LRGs}

\begin{figure}
{\epsfxsize=8.truecm
\epsfbox[18 144 592 718]{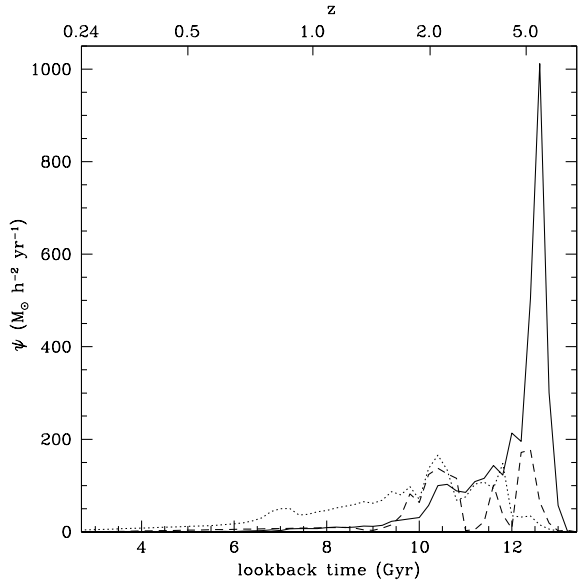}}
\caption{Three examples of star formation histories of $z=0.24$ LRGs extracted 
from the Bower et~al. model, plotted as a function of the lookback time from $z=0$. 
The upper axis gives the corresponding redshift. The star formation rate plotted 
is the sum of the star formation rate in {\it all} of the progenitor galaxies 
present at a given redshift. 
}
\label{fig:sfh}
\end{figure}

\begin{figure}
{\epsfxsize=8.truecm
\epsfbox[18 144 592 718]{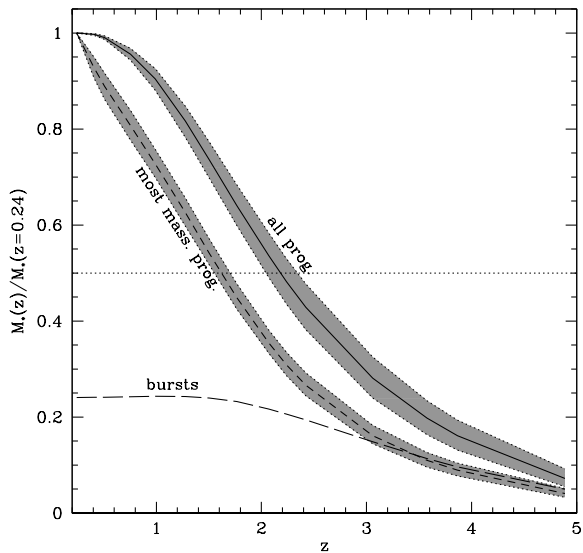}}
\caption{
The evolution of the total stellar mass of SDSS LRGs, as predicted by the 
Bower et~al. model. The stellar mass in place at a given redshift is expressed 
as a fraction of the mean stellar mass of SDSS LRGs at $z=0.24$. The mean fraction 
of mass in place as a function of lookback time (defined so that $t(z=0)=0$), 
summed over all progenitors and for the most massive progenitor of the SDSS LRGs, 
is shown by the solid and short-dashed lines respectively. 
The shaded region indicates the 1$\sigma$ scatter in 
this ratio. The long-dashed line shows the fraction of the mean mass accounted for by 
bursts of star formation, in all progenitors of the $z=0.24$.
}
\label{fig:sme}
\end{figure}

Semi-analytical galaxy formation models trace the full star formation 
and merger histories of galaxies. This allows us to build up a picture 
of how the stellar mass of LRGs was assembled and how the LRG population 
changed between the median redshifts of the 2SLAQ and SDSS surveys. 
The merging history, in particular, has implications for the clustering 
expected on small scales, which, as we saw in the previous section, is 
in excellent agreement with the observational estimates (e.g. Masjedi 
et~al. 2006). 

We first consider the star formation and mass assembly histories of LRGs. 
There are two ways in which a galaxy can acquire stellar mass in 
hierarchical models: 1) through the formation of new stars and 2) through 
the accretion of pre-existing stars in galaxy mergers (Baugh, Cole \& 
Frenk 1996; Kauffmann 1996). A nice discussion of the relative importance 
of these two processes for brightest cluster galaxies can be found in 
de Lucia \& Blaizot (2007). 

In Fig.~\ref{fig:sfh} we show some examples of star formation 
histories for three $z=0.24$ LRGs in the Bower et~al. model. 
The star formation rate plotted is the sum of the star formation rate 
over {\it all} of the progenitor galaxies at each redshift. The star formation 
history contains contributions from quiescent star formation in galactic 
disks and from starbursts triggered by galaxy mergers or dynamically 
unstable disks in the case of the Bower et~all. model. In general, 
the galaxy star formation histories predicted by hierarchical models 
tend to be more complex than the simple, one parameter, exponentially 
declining models typically considered in the literature (for some examples of 
star formation histories generated by semi-analytical models, see Baugh 2006). 
Fig.~\ref{fig:sfh} confirms that the LRG selection isolates a subset 
of galaxies in the model with more passive star formation histories, 
which are closer to exponential models (though these examples still 
display significant structure in their star formation histories at 
high redshift). The three examples have similar forms, with a peak 
at a lookback time $10 \la t \la 12$ Gyrs, followed by a smooth decay. 
In one of the examples, plotted with the dotted line, there is still 
appreciable star formation at $z=0.24$. Around 5\% of LRGs in the 
Bower et~al. model display star formation rates at $z=0.24$ 
in excess of $0.1\,{\rm M_{\odot}\,h^{-2}\,yr^{-1}}$, with the largest 
being $30\,{\rm M_{\odot}\,h^{-2}\,yr^{-1}}$. These low star formation 
rates indicate that for the bulk of LRGs in the model, ongoing star 
formation is not an important channel for increasing the stellar mass 
of LRGs, given the large stellar masses predicted for these galaxies. 

The evolution of the stellar mass of LRGs with redshift is shown in 
Fig.~\ref{fig:sme}. In this plot, we take LRGs of similar stellar mass 
at $z=0.24$ from the Bower et~al model and track the build up of their  
stellar mass with redshift.
The solid line shows how the mean stellar mass of the sample 
of LRGs builds up over time, expressed as a fraction of the mean mass 
of the LRGs at $z=0.24$. Here we sum over all the progenitors of SDSS 
LRGs. Half of the stellar mass of the $z=0.24$ LRGs was already in place 
at $z=2.2$. The mean change in stellar mass since $z=0.5$ is 
around $1\%$. If, instead, we consider only the most massive 
progenitor, the figure reveals that half of the mass was already in 
one object at $z=1.6$. Since $z=0.5$, the mean fractional change in 
stellar mass of the biggest progenitor is just over 0.1; since $z \sim 1$, 
the average stellar mass has grown by only 25\%. This evolution is 
in agreement with estimates inferred from the observed evolution of 
the luminosity function for a matched sample of LRGs (i.e. by 
considering the subset of 2SLAQ LRGs which have similar properties 
to the SDSS LRGs; Wake et~al. 2006). 
Fig.~\ref{fig:sme} also shows the contribution to the stellar mass of LRGs 
from bursts of star formation initiated by galaxy mergers or by the 
formation of bars in dynamically unstable discs. 
In total, this channel of star formation is only responsible for 
around $30\%$ of the stellar mass of SDSS LRGs. Furthermore, the level 
of contribution from this mode of star formation has changed little 
since a redshift of $z \sim 1.5$. This reflects the general trend for 
mergers to become more gas poor ( or ``dry'') with declining redshift 
in hierarchical models, due to the increasing consumption of gas by quiescent 
star formation in galactic discs, and the overall decline in the 
merger rate towards the present day.  Given the relatively small star 
formation rates predicted in model LRGs since $z \sim 1$, the steady 
increase in the stellar mass of LRGs over the redshift interval 
$z=1$ to $z=0.24$ is driven primarily by gas-poor galaxy mergers, 
which reassemble pre-formed stars. 

We now look in more detail at how SDSS LRGs build up their mass in 
galaxy mergers. The number of progenitor galaxies depends upon whether 
or not a mass cut is applied before a progenitor is counted. If a mass 
cut is not used, the number of progenitors obtained is likley to be dominated 
by low mass galaxies, which only bring in a small fraction of the galaxy's 
mass. Masjedi et~al. (2006) used their measurement of the correlation 
function on small scales to constrain a simple model for mergers between 
LRGs, prior to the median redshift of the SDSS sample. Here we do not 
attempt to consider only those progenitors which are matched to the SDSS 
sample selection. Instead, we consider progenitors at $z=0.5$ which 
account for 30\% of more of the stellar mass of the SDSS LRG at $z=0.24$. 
We then further distinguish between progenitors which satisfy the 2SLAQ 
selection and those that do not. 

Table~\ref{tab:nprogs} shows the number of progenitors of SDSS 
LRGs predicted by the Bower et~al model. 
The first row gives the percentage of SDSS LRGs which have 0,1,2 or 3 
progenitors at $z=0.5$ which each account for 30\% or more of the mass 
of the $z=0.24$ LRG. Typically, in the model, SDSS LRGs have one such 
progenitor at $z=0.5$. Only 11\% of LRGs have more than one progenitor 
which represents 30\% or more of the mass. In the case where a SDSS LRG has 
only one progenitor with $\ge 30\%$ of the mass, then in 70\% of cases this 
progenitor will be a 2SLAQ LRG. 
When there are two sizeable progenitors present at $z=0.5$, 
then in half of the cases, one galaxy is a 2SLAQ LRG and the other progenitor 
fails to meet the 2SLAQ LRG definition. In around one third of cases, both 
progenitors are LRGs. 

\begin{table}
\begin{center}
 \begin{tabular}{ccccc}
 \hline
  \# progenitors $>30\%$ mass& 0& 1 & 2 & 3 \\
 \hline 
     (\%)            & 0.3 & 88.6 & 11.0 & 0.1 \\
 \hline 
  \# progenitors $>30\%$ mass &  &   &   &  \\
     and 2SLAQ LRG            &  &   &   &  \\ 
 \hline
  0 & 100 & 27 & 10 & 29 \\
  1 &  -  & 73 & 55 & 29 \\
  2 &  -  &  - & 35 & 29 \\
  3 &  -  &  - & -  & 13 \\
 \hline
 \end{tabular}
  \caption{
The nature of the progenitors of SDSS LRGs in the Bower et~al. (2006) 
model. The progenitor galaxies are identified at $z=0.5$ and by definition 
are required to account for at least 30\% of the stellar mass of the 
LRG at $z=0.24$, the median redshift of the SDSS sample.
The second row gives the percentage of SDSS LRGs with 0, 1, 2 and 3 
such progenitors. The rows give the percentage of cases in which 
a galaxy has the stated number of progenitors which themselves 
satisfy the LRG selection criteria at $z=0.5$. 
}
 \label{tab:nprogs}
 \end{center}
\end{table}

\section{Discussion and conclusions} 

In this paper we have extended the tests of hierarchical galaxy 
formation models to include predictions for the properties of a special 
subset of the galaxy population called luminous red galaxies (LRGs). 
Given their rarity, bright luminosities and extreme colours, LRGs 
represent a stern challenge for the models. They are particularly 
interesting from the point of view of developing the model physics, 
since the abundance and nature of LRGs probes precisely the regime 
in which the models are currently most uncertain, the formation of 
massive galaxies. Historically, hierarchical models have tended to 
overproduce bright galaxies at the present day (see Baugh 2006). 
The phenomena invoked to restrict the growth of large galaxies locally, 
naturally, have an impact on the form of the bright end of the galaxy 
luminosity function at intermediate and high redshifts, where 
LRGs dominate. In addition, LRGs have the red colours expected of a 
passively evolving stellar population, which restricts the range of 
possible star formation histories for these galaxies. 

It may appear odd to talk about producing model 
predictions after an observational dataset has been constructed. The 
models considered in this paper contain parameters whose values were 
fixed by requiring them to reproduce a subset of the data available 
for the local galaxy population (for a discussion see Cole et~al. 2000). 
None of the datasets used for this purpose make explicit reference 
to the redshifts of interest for the LRG surveys discussed here, nor 
were red galaxies singled out for special attention in the process 
of setting the model parameters. We do, however, require that our 
semi-analytical models reproduce as closely as possible the bright 
end of the present day luminosity function of all galaxies, which 
does tend to be dominated by red galaxies with passive stellar 
populations (e.g. Norberg et~al. 2002). Therefore, by comparing the 
models to the observed properties of the LRG population, we are in effect 
testing the physics which govern the evolution of the bright 
end of the luminosity function, as traced by objects with the special 
colours of LRGs. 

The two models considered, Baugh et~al. (2005) and Bower et~al. (2006), 
enjoy a considerable number of successes and, inevitably, have some 
shortcomings (see Section 2). It is important to be clear that in this 
paper, we have not adjusted or tinkered with {\it any} of the parameters 
of the published models in order to improve the comparison of the model 
output with the observational data. This ``warts and all'' exercise 
illustrates the appeal of the semi-analytical approach, in that a given 
model yields a broad range of outputs which are directly testable 
against observations. In both models, LRGs are predominantly bulge 
dominated galaxies (although 20-40\% are expected to be spirals with 
old stellar populations), with 
velocity dispersions of $\sigma \sim 320 {\rm km s}^{-1}$ and stellar masses 
around $1-2 \times 10^{11} h^{-1}M_{\odot}$, which is higher than observed. 
The models give different 
predictions for the radii of LRGs, with the Baugh et~al model predicting  
the larger LRGs. Both models fail to produce bright spheroids that are 
large enough to match the locally observed radius-luminosity relation 
(see Almeida et~al. 2007 for a discussion of how the sizes of spheroids 
are computed in the models and for possible solutions to this problem). 

The Baugh et~al. and Bower et~al. models are two feasible simulations of the 
galaxy formation process, which differ in several ways, as we reviewed in 
Section 2 (see also the comparison in Almeida et~al. 2007). A key difference 
between the models, in terms of the analysis presented in this paper, is 
the form of the physics invoked to quench the formation of massive galaxies. 
In both models the amount of gas cooling from the hot halo, to provide the 
raw material for star formation, is reduced by quite different means. 
Baugh et~al. invoke a wind which expels cold gas from intermediate mass 
haloes. This gas is assumed to be ejected with such vigour that it does 
not get recaptured by more massive haloes in the merger hierarchy. Hence, 
in this model, the more massive haloes contain fewer baryons 
than expected from the universal baryon fraction, and therefore less gas is 
available to cool from the hot halo. One controversial aspect of this scheme  
is the energy source required to drive the wind. Benson et~al. (2003) showed 
that the energy produced by supernova explosions is unlikely to be sufficient 
to power a wind of the strength required to reproduce the sharpness of the 
break in the local galaxy luminosity function, and argued 
that the accretion of gas onto a central 
supermassive black hole could be the solution. Bower et~al. invoked an AGN 
feedback model in which the luminosity of the AGN heats the hot halo (see 
also Croton et~al. 2006; see Granato et~al. 2004 for an 
alternative model). This suppresses the cooling flow in massive haloes which 
have quasi-static hot gas atmospheres. 

The predictions of the Baugh et~al. and Bower et~al. models bracket the 
observed luminosity function of LRGs, with the Bower et~al. model giving  
the better overall agreement with the SDSS and 2SLAQ results. 
The shape and normalization of the z=0.24 LRG luminosity function predicted 
by the Bower et~al. model are in excellent agreement with the observations. 
This is remarkable when one bears in mind that LRGs are an order of 
magnitude less common than $L_*$ galaxies. The Baugh et~al. model on the 
other hand, whilst predicting a similar number density of LRGs, gives a 
poor match to the shape of the observed luminosity function. 
At $z=0.5$, the agreement is less good, with the predictions only coming 
within a factor of 2 of the observed abundance. This implies that the models 
may not be tracking the evolution of the bright end of the luminosity 
function accurately over such a large lookback time (40\% of the age of the 
universe), at least for galaxies matching the 2SLAQ selection. Whilst this 
discrepancy suggests that there are problems modelling the evolution 
of the red galaxy luminosity function, it is important to note that 
the Bower et~al. model does give a good match to the inferred evolution of 
the stellar mass function, to much higher redshifts than that of the 
2SLAQ sample. We investigated whether it was possible to tune the predictions 
of the Baugh et~al. model to better match the LRG luminosity function; this 
exercise proved to be unsuccessful suggesting that a more substantial 
revision to the ingredients of this model, involving further suppression 
of gas cooling in massive haloes, is required. 

Semi-analytical models predict the star formation histories of 
galaxies, based upon the mass of cold gas which accumulates through 
cooling and galaxy mergers, and a prescription for computing an
instantaneous star formation timescale 
(examples of star formation histories extracted 
from the models are given in Baugh 2006). 
As expected, the stellar populations of model LRGs are old, 
with luminosity weighted ages in the region of 4-8 Gyr for the SDSS 
selection, with the Bower et~al. model returning the more elderly stars 
(similar results were reported by de Lucia et~al. 2006 and Croton et~al. 2006 
for massive elliptical galaxies). The semi-analytical model can track 
the build-up of the stellar mass of LRGs, considering all of the progenitor 
galaxies. There is little recent star formation in any 
of the progenitor galaxies of SDSS LRGs; averaging over all progenitors, 
typically 50\% of the $z=0.24$ stellar mass of the LRG has already formed by 
a redshift of $z \sim 2.2$. However the mass of the main progenitor branch 
is still growing over this redshift interval. Around half of the mass 
in the biggest progenitor is put in place since $z \sim 1.5$ through 
galaxy mergers of ready-made stellar fragments (for a discussion of the  
difference between the formation time of the stars and the assembly 
time of the stellar mass, see de Lucia \& Blaizot 2007). On average, 
only 25\% of the stellar mass of the LRG is added after $z \sim 1$, in 
line with observational estimates of the evolution of the 
stellar mass function, which indicate that many of the most massive 
galaxies are already in place by $z \sim 1$ (e.g. Bauer et~al. 2005; 
Bundy, Ellis \& Conselice 2005; Wake et~al. 2006).

Perhaps the most spectacularly successful model prediction is for the 
clustering of LRGs. Masjedi et~al. (2006) estimated the two-point 
correlation function of SDSS LRGs in real space, free from the 
distortions in the clustering pattern induced by the peculiar motions 
of galaxies. These authors found that the real-space correlation 
function of LRGs is a power law over three and a half decades in pair 
separation, down to scales of $r \approx 0.01 h^{-1}{\rm Mpc}$. 
Masjedi et~al. argued that current halo occupation distribution models 
could not reproduce such a steep correlation function on small scales 
because these models asume that galaxies trace the density profile of the 
dark matter halo, which is shallower than the observed correlation 
function. This line of reasoning is spurious, as HODs can produce 
realizations of the two-point correlation function with different 
small-scale slopes for different galaxy samples, even when the different 
samples trace the dark matter (see, for example, Figure 22 of Berlind et~al. 2003 
which compares the correlation functions of old and young galaxies).
The small scale slope depends on the interplay between two factors: 
the number of galaxies within a dark matter halo and the range of halo 
masses which contain more than one galaxy (e.g. Benson et~al. 2000). 
The Bower et~al. model can readily produce predictions of galaxy 
clustering down to such small scales since it is embedded in the 
Millennium simulation (Springel et~al. 2005). The correlation function 
predicted by the Bower et~al. model agrees impressively well with the 
observational estimate by Masjedi et~al. The HOD used by Masjedi et~al. 
is actually a poor description of the HOD predicted in the Bower 
et~al. model. Further support for the number of LRGs predicted as 
a function of halo mass comes from the degree of damping of the 
correlation function seen on small scales in redshift-space. 
The virialized motions of LRGs within a common halo gives a contribution 
to the peculiar velocity of these galaxies, which results in the structure 
appearing stretched when the distance to the LRG is 
inferred from its redshift. 
This damping would not be apparent in the case of a maximum of 
one LRG per halo. 

Overall, the agreement between the model predictions and the 
observation of LRGs is encouraging, demonstrating the true 
predictive power of semi-analytical models. The two models we 
have tested have quite different mechanisms to regulate the formation 
of massive galaxies, with the Bower et~al. model invoking ``AGN feedback'' 
and the Baugh et~al. model relying on a ``superwind''; in the former, the 
raw material for star formation is prevented from cooling in the first place 
in massive haloes, whilst in the latter cold gas is expelled from the halo 
before it can form stars. The Bower et~al. model does the best in terms of 
matching the abundance of LRGs, particularly at $z=0.24$. This success 
is repeated for extremely red objects (EROs) at higher redshifts than 
the samples considered here, as presented by Gonz\'alez-P\'erez et~al. 
(in preparation). The Baugh et~al. model does less well at reproducing 
the number of LRGs and EROs.
Whilst problems remain in predicting the radii of spheroids and the 
precise evolution of LRG luminosity function, it is clear that these 
objects can be accommodated in hierarchical models.

\subsection*{ACKNOWLEDGEMENTS}
{\small
C. A. gratefully acknowledges support in the form of a scholarship 
from the Science and Technology Foundation (FCT), Portugal. CMB is 
supported by the Royal Society. AJB acknowledges support from the 
Gordon and Betty Moore Foundation. This work was supported in part 
by a rolling grant from PPARC. 
We thank the referee for providing a detailed and helpful report. 
We acknowledge comments and suggestions from Bob Nichol, Nic Ross 
and Donald Schneider, and the contributions of Shaun Cole, Carlos Frenk, 
John Helly and Rowena Malbon to the development of the {\tt GALFORM} code.
}

\end{document}